\documentclass{jfm}
\usepackage{graphicx}
\usepackage{epstopdf, epsfig, color, amsmath, amsfonts}
\usepackage{subfig}
\usepackage{pifont}
\usepackage{bm}
\usepackage{tikz}
\tikzstyle{long dashdotted}=[dash pattern=on 6pt off 3pt on \the\pgflinewidth off 3pt]
\graphicspath{{Figures/}}

\definecolor{C0}{RGB}{31,119,180}
\definecolor{C1}{RGB}{255,127,14}
\definecolor{C2}{RGB}{44,160,44}
\definecolor{C3}{RGB}{214,39,40}
\definecolor{C4}{RGB}{148,103,189}

\shorttitle{Grain size segregation during bedload transport on steep slopes}
\shortauthor{R. Chassagne, R. Maurin, J. Chauchat, JMNT Gray and P. Frey}

\title{Grain size segregation during bedload transport on steep slopes}

\author{R\'emi Chassagne\aff{1}
  \corresp{\email{remi.chassagne@irstea.fr}},
  Rapha\"el Maurin\aff{3},
  Julien Chauchat\aff{2},
  J. M. N. T. Gray\aff{4}
 \and Philippe Frey\aff{1}}

\affiliation{\aff{1}Univ. Grenoble Alpes, Irstea, UR ETNA, 38000 Grenoble, France
\aff{2}Univ. Grenoble Alpes, LEGI, CNRS UMR 5519 - Grenoble, France
\aff{3}IMFT, Univ. Toulouse, CNRS - Toulouse, France
\aff{4}Department of Mathematics, University of Manchester, Manchester M13 9PL, UK}

\begin{document}

\maketitle

\begin{abstract}
Size segregation in bedload transport is studied numerically with a coupled fluid-discrete element model. Starting from an initial deposit of small spherical particles on top of a large particle bed, the segregation dynamics of the bed is studied as it is driven by the fluid flow. Focusing on the quasi-static part of the bed, the small particles are observed to segregate as a layer of constant thickness at a velocity constant in time and independent of the number of small particles. The segregation velocity is observed to be directly linked to the inertial number at the bottom of the layer, and to increase linearly with the size ratio. 
While the macroscopic behavior is independent of the concentration in small particles, an analysis in the framework of the continuous model of \cite{thornton2006} shows that the dynamics results from an equilibrium between the influence of the local concentration and the inertial number forcing. Deriving an analytical solution of the continuous model, it is shown that the diffusion coefficient should have the same dependency on the inertial number as the segregation flux. Implementing the segregation and the diffusive fluxes in the continuous model, it is shown that they quantitatively reproduce the discrete simulations. 
These results improve the understanding of the size segregation dynamics and represent a step forward in the upscaling process of polydisperse granular flow in the context of turbulent bedload transport.  
\end{abstract}

\section{Introduction}\label{sec:intro}
    Bedload transport, the coarser sediment load transported by a flowing fluid in contact with the mobile stream bed by rolling, sliding and/or saltating, has major consequences for public safety, water resources and environmental sustainability. In mountains, steep slopes drive an intense transport of a wide range of grain sizes implying size sorting or segregation, which is largely responsible for our limited ability to predict sediment flux and river morphology \citep{bathurst2007, frey2011, dudill2018}.

The present study focuses on size segregation. For large size ratios, the small particles can percolate without external forcing by gravity into the bed of larger particles \citep{bridgwater1971, dudill2017}. For the smaller size ratios, percolation is not possible without deformation of the bed, which can be achieved by shearing or vibrating. This dynamic segregation results from the combination of kinetic sieving \citep{middleton1970} and squeeze expulsion \citep{savage1988}. Kinetic sieving is based on the idea that, when sheared, granular media experience velocity fluctuations creating holes between particles, in which small particles are more probable to fall into than large particles. In opposition, squeeze expulsion tends to push all particles upwards with the same probability. The combination of both processes, which for brevity will just be called kinetic sieving in the following, results in a net flux downward for the small particles and upward for the large, leading to an inverse graded bed. Kinetic sieving has then been studied experimentally and numerically in multiple configurations such as dry granular avalanches \citep{savage1988, dolgunin1995, wiederseiner2011, fan2014, jones2018}, shear cells \citep{golick2009, may2010, vandervaart2015, fry2018}, annular rotating drum \citep{gray2011} and laminar bedload configurations \citep{ferdowsi2017}.\\

In order to identify the mechanisms and comment on the literature on kinetic sieving, a dimensional analysis is performed. This analysis is made in the dry limit case, consistent with the results of \cite{maurin2016} showing that the fluid just acts as a forcing mechanism in turbulent bedload transport. The different parameters of the problem are
\begin{equation}
 \begin{array}{cccccccccc}
 d_s, & d_l, & \phi_s, & \rho^p, & g, & \dot{\gamma}^p, & P, & w_s,
 \end{array}
\end{equation}
where $d_s$ (resp. $d_l$) is the diameter of small (resp. large) particles, $\phi_s$ is the local concentration in small particles, $\rho^p$ is the particle density,  $g=9.81ms^{-2}$ is the gravity acceleration, $\dot{\gamma}^p$ is the particle shear rate, $P$ is the granular pressure and $w_s$ is the segregation velocity of the small particles. These eight variables involve only three units (length, time and mass), so the problem depends on five dimensionless parameters. Choosing $d_l$, $\sqrt{d_l/g}$ and $\rho^pd_l^3$ as reference length, time and mass units respectively, the following relation is obtained
\begin{equation}
 \dfrac{w_s}{\sqrt{gd_l}} = \mathcal{G}\left(r, \phi_s, \sqrt{\frac{d_l}{g}}\dot{\gamma}^p, \frac{P}{\rho^pd_lg}\right),
 \label{eq:piTh}
\end{equation}
where $r=d_l/d_s$ is the size ratio between large and small particles and $\mathcal{G}$ is an unknown function. It is assumed that relation~\eqref{eq:piTh} can be written in a power law form,
\begin{equation}
 \dfrac{w_s}{\sqrt{gd_l}} \propto (r-1)^l\left(\sqrt{\frac{d_l}{g}}\dot{\gamma}^p\right)^{m}\left(\frac{P}{\rho^pd_lg}\right)^{n}\mathcal{G}_1(\phi_s),
 \label{eq:piTh2}
\end{equation}
where the dependence on the size ratio has been written as $(r-1)^l$ in order to cancel segregation in the monodisperse limit. \\

In the theory of \cite{savage1988}, $\dot{\gamma}^p$ is predicted to be a controlling parameter for size segregation as the shear rate allows particles to move relatively faster than the one below them and find holes in which to segregate. With experiments of bidisperse flows down inclined planes, \cite{savage1988} measured concentration profiles at different streamwise positions and found good agreement between the theory and experiments. With discrete element method (DEM) simulations of dry bidisperse mixtures in a heap flow configuration, \cite{fan2014} measured the percolation velocity of each species and found a linear relation between the percolation velocity and the shear rate.
In a shear cell experiment with a rotating bottom plate and a weighthed top plate, \cite{golick2009} evidenced the effect of pressure on segregation. Small particles were initially placed above large particles, and they characterized the segregation rate by measuring the volume evolution of the cell. They found that the segregation rate slows down strongly with granular pressure. In a dry confined shear cell, \cite{fry2018} also observed this effect of pressure on segregation and found a power law relating the segregation rate and the inertial number $I$ \citep{gdrmidi2004}. The latter  scaling represents an extension of the \cite{savage1988} theory in order to take into account both the effect of the shear rate and pressure on segregation. It corresponds in~\eqref{eq:piTh2} to $n=-m/2$, which can be rewritten as
\begin{equation}
 \dfrac{w_s}{\sqrt{gd_l}} \propto (r-1)^lI^{m}\mathcal{G}_1(\phi_s),
 \label{eq:piTh3}
\end{equation}
with $I$ the large particle inertial number defined as
\begin{equation}
 I = \dfrac{d_l\dot{\gamma}^p}{\sqrt{P/\rho^p}}
\end{equation}

suggesting that the segregation velocity is a function of the inertial number, the size ratio and the local concentration. 

The size ratio between large and small particles $r$ is also a key parameter for size segregation. Smaller particles are expected to segregate more easily. In the annular cell configuration, \cite{golick2009} showed that the segregation rate is an increasing function of the size ratio with a maximum around $r=2$. In 2D DEM simulations of sheared and confined granular flow, \cite{guillard2016} measured the segregation force applied on a single large particle placed in a small particle mixture. They also showed that the segregation force increases with $r$ and observed the same maximum around $r=2$.

The effect of the small particle concentration $\phi_s$ has also been widely studied, and different forms for $\mathcal{G}_1$ have been proposed \citep{bridgwater1985, savage1988, dolgunin1995, gray2005, may2010, gajjar2014, vandervaart2015, fan2014}. The small particles have been shown to segregate less easily if they are more concentrated and the downward velocity should vanish for a pure phase of small particles. For this purpose, the simplest form $\mathcal{G}_1(\phi_s)=1-\phi_s$, has been proposed by \cite{dolgunin1995}. In oscillating cell experiments, \cite{vandervaart2015} measured the time necessary to achieve complete segregation with an initial mixture ranging from one small particle falling into a bed of large particles to a unique large particle going up in a small particle bed. They observed that both extreme cases were not symmetric, resulting in an asymmetry of the vertical velocity with the concentration and \cite{gajjar2014} proposed thereby a quadratic form for $\mathcal{G}_1$ with $\phi_s$. Simulations by \cite{fan2014, jones2018} of dry bidisperse avalanches, observed this non linearity of the velocity when $\phi_s$ is close to zero, but concluded that the form of \cite{dolgunin1995} is at first order a good approximation.\\

The literature review on dry granular flows underlines the qualitative understanding of size-segregation for simple flows. Bedload transport can be seen as a granular flow, where the coupling with the fluid induces strong gradients in the vertical direction and a complex forcing, which could challenge the classical picture of segregation. Few studies have been made on kinetic sieving in bedload transport and more remains to be done for a clear understanding of the processes at play. \cite{ferdowsi2017} studied experimentally size segregation in laminar bedload transport and performed dry granular flow simulations. They studied the formation of armour, i.e. the segregation of large particles to the top, starting from a mixture of small and large particles. They showed that the process seems to be a granular phenomenon and reproduced their experimental results in the framework of continuous segregation modelling, using an ad-hoc parametrisation of the model. \cite{hergault2010} and Frey et al. (2019) studied size-segregation in turbulent bedload transport, considering a quasi-2D channel enabling particle tracking. They found that the particles move down as a layer into the bed, and related the segregation velocity to the granular shear rate.\\

Modelling turbulent bedload transport segregation phenomenon at the particle scale is computationally demanding and can be only achieved for very small domains. In this context, one of the main goals for segregation in turbulent bedload transport is to be able to do upscaling to the framework of continuous modelling. In this case, it includes the modelling of the fluid phase, the granular phase and the evolution of the different classes of particles with respect to one another. Such a segregation model has been developed by \cite{gray2005}, \cite{thornton2006} and \cite{gray2006}. This three phase model is based on the assumption that the overburden granular pressure is not shared equally between large and small particles \citep{gray2005}. Substituting this into the momentum conservation equation of each constituent and placing the problem within the framework of the mixture theory, an evolution equation for the phase of small particles is found
\begin{equation}
 \dfrac{\partial \phi_s}{\partial t} + \nabla \cdot \left(\phi_s\bm{u}\right) - \dfrac{\partial F_s}{\partial z} = \nabla \cdot \left(D \nabla \phi_s\right),
 \label{eq:gray_init}
\end{equation}
where $\bm{u}$ is the bulk velocity field, $D$ the diffusion coefficient, $F_s$ the segregation flux. The physics relies on the expression of the segregation flux and of the diffusion flux. The former is defined as $F_s = \phi_s w_s$ and can therefore directly be linked to the above stated results. Introducing the dimensional analysis~\eqref{eq:piTh3} and considering the simplest form of \cite{dolgunin1995} for $\mathcal{G}_1$ the following segregation flux is obtained
\begin{equation}
 F_s \propto (r-1)^lI^{m}\phi_s(1-\phi_s).
 \label{eq:flux_init}
\end{equation}
In granular flows, diffusion has been mainly studied in the case of self-diffusion. By analogy to the thermal diffusion of molecules, \cite{campbell1997} tracked, in numerical simulations of dilute monodisperse sheared flows, the random motions of particles and showed that the diffusion coefficient should scale with the shear rate. In dense granular flows, \cite{utter2004} also observed, in 2D Couette flow experiments, the dependency of the diffusion coefficient with the shear rate. The diffusion mechanism considered in this paper is the mixing of one class of particles into another. This kind of diffusion is expected to be related to self-diffusion, but to the best of our knowledge, no study has been done in order to determine the physical mechanism controlling this diffusion.\\

While the literature review underlined the progress in the understanding and modelling of kinetic sieving, more physically-based parametrisations of the continuous model are still lacking, in particular in the complex turbulent bedload transport configuration. In the present contribution, size segregation in turbulent bedload transport is studied numerically considering fluid-DEM simulations. The aim is to understand size segregation in the quasi-static part of the flow and to improve the continuous modelling parametrisations. In particular the influence of the three parameters $r$, $I$ and $\phi_s$ and the values of exponents $l$ and $m$, will be investigated through numerical experiments in the bedload configuration.

After presenting the numerical model (\S~\ref{sec:model}), numerical simulations of turbulent bedload transport will be presented and analysed based on the different expected dependencies from the dimensional analysis (\S~\ref{sec:dem}). The results are then studied, through an analytical analysis, in the framework of continuous modelling and highlight the local segregation mechanisms. This analysis shows that diffusion and segregation should have the same dependence on the inertial number and enables quantitatively reproducing the DEM results (\S~\ref{sec:mechanism}). 

\section{Fluid-DEM model for bidisperse system}\label{sec:model}
    The numerical model used to simulate size-segregation in turbulent bedload transport is a three-dimensional Discrete Element Method (DEM) using the open-source code YADE \citep{smilaueretal.2015} coupled with a vertical one-dimensional turbulent fluid model. This code has been validated with particle-scale experiments \citep{frey2014} in \cite{maurin2015} for mono-disperse situations. A brief summary of the model formulation as well as a description of its adaptation to bi-disperse situations is now given.

\subsection{Granular phase}

DEM is a Lagrangian method based on the resolution of contacts between particles. For each spherical particle $p$, the motion of the particle is obtained from Newton's second law, and a momentum equation and an angular momentum conservation equation are solved
\begin{align}
 m^p\dfrac{d^2 \bm{x}^p}{dt^2} = \bm{f}_g^p + \bm{f}_c^p + \bm{f}_f^p,\\
 \mathcal{I}^p\dfrac{d\bm{\omega}^p}{dt} = \bm{\mathcal{T}} = \bm{x}_c \times \bm{f}_c^p,
\end{align}
where $m^p$, $\bm{x}^p$, $\bm{\omega}^p$ and $\mathcal{I}^p$ are respectively the mass, position, angular momentum and moment of inertia of particle $p$. The three major forces are: $\bm{f}_g^p$ the gravitational force, $\bm{f}_c^p$ the inter-particles contact forces and $\bm{f}_f^p$ the interaction forces with the fluid. The inter-particles contact forces are classically defined as a spring-dashpot system \citep{schwager2007} composed of a spring of stiffness $k_n$ in parallel with a viscous damper of coefficient $c_n$ in the normal direction; and a spring of stiffness $k_s$ associated with a slider of friction coefficient $\mu$ in the tangential direction. The normal and tangential contact forces reads accordingly
\begin{equation}
 \begin{array}{rcl}
  F_n &=& -k_n\delta_n - c_n\dot{\delta}_n,\\
  F_t &=& -min(k_s\delta_t, \mu_pF_n),\\
 \end{array}
\end{equation}
where $\delta_n$ (resp. $\delta_t$) is the overlap between particles in the normal (resp. tangential) direction. For each class of particles, the value of $k_n$ and $k_s$ are computed in order to stay in the rigid limit of grains \citep{roux2002, maurin2015}. The normal stiffness in parallel with the viscous damper, defines a restitution coefficient representative of the loss of energy during  collisions, which is fixed to $e_n=0.5$ for each contact. Finally the friction coefficient is fixed to $\mu = \arctan(0.4)$ for all particles independently of their diameter.

The third type of forces concerns the interactions with the fluid which are restricted in this model to the buoyancy force and the drag force \citep{maurin2015}. They are defined as:
\begin{equation}
 \bold{f}_b^p = -\dfrac{\pi d^{p3}}{6}\nabla P^f,
\end{equation}
\begin{equation}
 \bold{f}_D^p = \dfrac{1}{2} \rho^f \dfrac{\pi d^{p2}}{4}C_D ||\left<\bold{u}\right>_{\bold{x}^p}^f-\bold{v}^p||\left( \left<\bold{u}\right>_{\bold{x}^p}^f-\bold{v}^p \right),
 \label{eq:drag}
\end{equation}
where $d^p$ denotes the diameter of particle $p$, $\left<\bold{u}\right>_{\bold{x}^p}^f$ is the mean fluid velocity at the position of particle $p$ and $\bold{v}^p$ is the velocity of particle $p$. The drag coefficient takes into account hindrance effects \citep{richardson1954} as $C_D = (0.4+24.4/Re_p)(1-\Phi)^{-3.1}$, with $\Phi$ the total solid phase (small and large particles) volume fraction and $Re_p = ||\left<\bold{u}\right>_{\bold{x}^p}^f-\bold{v}^p||d^p/\nu^f$ the particle Reynolds number.

\subsection{Fluid model}

The fluid model is a one dimensional vertical turbulent model inspired from the Euler-Euler model proposed by \cite{chauchat2018} and solves directly the volume-averaged momentum balance for the fluid phase,
\begin{equation}
 \rho^f(1-\Phi)\dfrac{\partial \left<u_x\right>^f}{\partial t} = \dfrac{\partial S_{xz}}{\partial z} - \dfrac{\partial R_{xz}}{\partial z} + \rho^f(1-\Phi)g_x -n\left<f_{f_x}^p\right>^s,
\end{equation}
where $\rho^f$ is the density of the fluid, $S_{xz}$ is the effective fluid viscous shear stress of a Newtonian fluid, $R_{xz}$ is the turbulent fluid shear stress and $n\left<f_{f_x}^p\right>^p$ represents the momentum transfer associated with the interaction forces between fluid and particles.

The viscous shear stress $S_{xz}$ is taken as
\begin{equation}
 S_{xz} = \rho^f(1-\Phi)\nu^f\dfrac{\partial \left<u_x\right>^f}{\partial z},
\end{equation}
with $\nu^f$ the pure fluid viscosity.
The Reynolds shear stress $R_{xz}$ is based on an eddy viscosity concept,
\begin{equation}
 R_{xz} = \rho^f(1-\Phi)\nu_t\dfrac{\partial \left<u_x\right>^f}{\partial z}.
\end{equation}
The turbulent viscosity $\nu_t$ follows a mixing length approach that depends on the integral of the solid volume fraction profile to account for the presence of particles \citep{li1995},
\begin{equation}
\begin{array}{lr}
 \nu_t = l_m^2|\dfrac{\partial \left<u_x\right>^f}{\partial z}|, & l_m(z) = \kappa \displaystyle{\int_0^z}\dfrac{\Phi_{max}-\Phi(\zeta)}{\Phi_{max}}d\zeta,
 \end{array}
\end{equation}
with $\kappa=0.41$ the Von-Karman constant and $\Phi_{max} = 0.61$ the maximal packing of the granular medium (random close packing).

The total momentum $n\left<f_{f_x}^p\right>^s$ transmitted by the fluid to the particles is computed as the sum over each class of the horizontal solid-phase average (denoted $\left<.\right>^s$) of the momentum transmitted by the drag force to each particle :
\begin{equation}
 n\left<f_{f_x}^p\right>^{s} = \dfrac{\Phi_s}{\pi d^{3}_s/6} \left<f_{f_x}^{p_s}\right>^s +  \dfrac{\Phi_l}{\pi d^{3}_l/6} \left<f_{f_x}^{p_l}\right>^s,
\end{equation}
and introducing the expression of the drag force \eqref{eq:drag}, 
 \begin{multline}
 n\left<f_{f_x}^p\right>^{s} = \dfrac{3}{4} \dfrac{\Phi_s \rho^f}{d_s}\left< C_D||\left<\bold{u}\right>_{\bold{x}^{p_s}}^f-\bold{v}^{p_s}||\left( \left<\bold{u}\right>_{\bold{x}^{p_s}}^f-\bold{v}^{p_s} \right)  \right>^s \\
 + \dfrac{3}{4} \dfrac{\Phi_l \rho^f}{d_l}\left< C_D||\left<\bold{u}\right>_{\bold{x}^{p_l}}^f-\bold{v}^{p_l}||\left( \left<\bold{u}\right>_{\bold{x}^{p_l}}^f-\bold{v}^{p_l} \right)  \right>^s,
\end{multline}
where $p_s$ (resp. $p_l$) denotes the ensemble of the small (resp. large) particles, $\Phi_s$ (resp. $\Phi_l$) the volume fraction of small (resp. large) particles, $d_s$ (resp. $d_l$) the diameter of small (resp. large) particles.

The solid phase volume average $\left<.\right>^s$ is defined following \cite{jackson2000} for which a cuboid weighting function $\mathcal{G}$ with the same length and width as the 3D domain is applied. In the vertical direction, in order to capture the strong vertical gradient of the mean flow, the vertical thickness $l_z$ of the box is chosen as $l_z = d_s/30$. This choice of weighting function has been validated by comparison with mono-disperse turbulent bedload transport experiments in \cite{maurin2015}. Therefore the mean values at position $z$ are computed as
\begin{align}
 \Phi_s(z) = \sum_{p_s}\int_{V_{p_s}} \mathcal{G}(|z-z^{\prime}|)dz^{\prime},\\
 \Phi_l(z) = \sum_{p_l}\int_{V_{p_l}} \mathcal{G}(|z-z^{\prime}|)dz^{\prime},
 \end{align}
\begin{equation}
 \left<\gamma\right>^s = \dfrac{1}{\Phi_s(z)}\sum_{p_s}\int_{V_{p_s}}\gamma(z^{\prime}) \mathcal{G}(|z-z^{\prime}|)dz^{\prime} 
 + \dfrac{1}{\Phi_l(z)}\sum_{p_l}\int_{V_{p_l}}\gamma(z^{\prime}) \mathcal{G}(|z-z^{\prime}|)dz^{\prime},
 \end{equation}
where $\gamma$ is a scalar quantity.

In this paper, this model is used for a bi-disperse situation to study size segregation in bedload sediment transport.

\section{DEM simulations of segregation dynamics}\label{sec:dem}
    In this section, the infiltration of fine particles into the bed made of larger particles is studied during turbulent bedload transport using the numerical model described in the previous section.

\subsection{Numerical setup}

\begin{figure}
\centering
 \subfloat[]{\includegraphics[height=5cm]{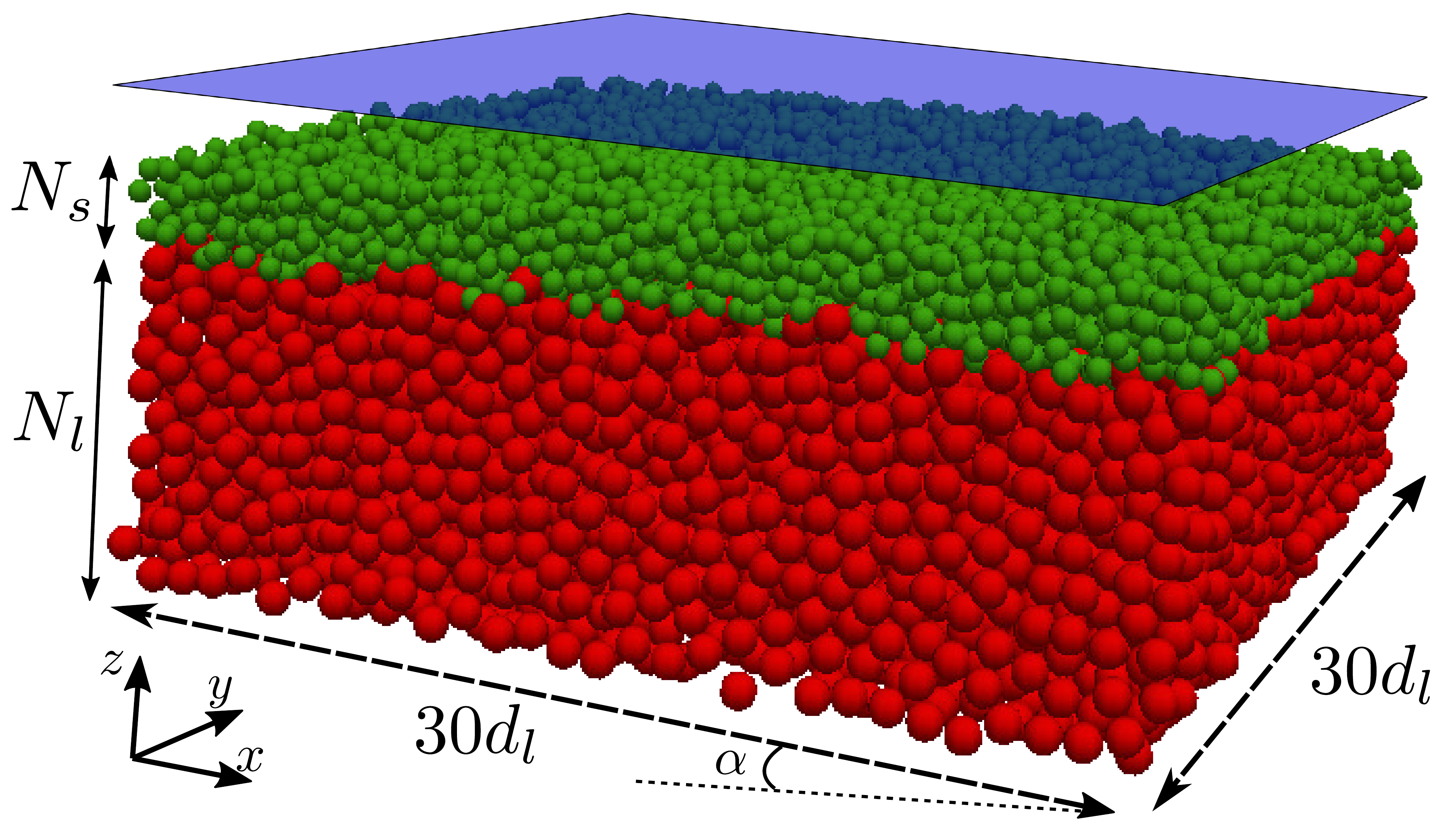}}\\
 \subfloat[]{\includegraphics[height=5cm]{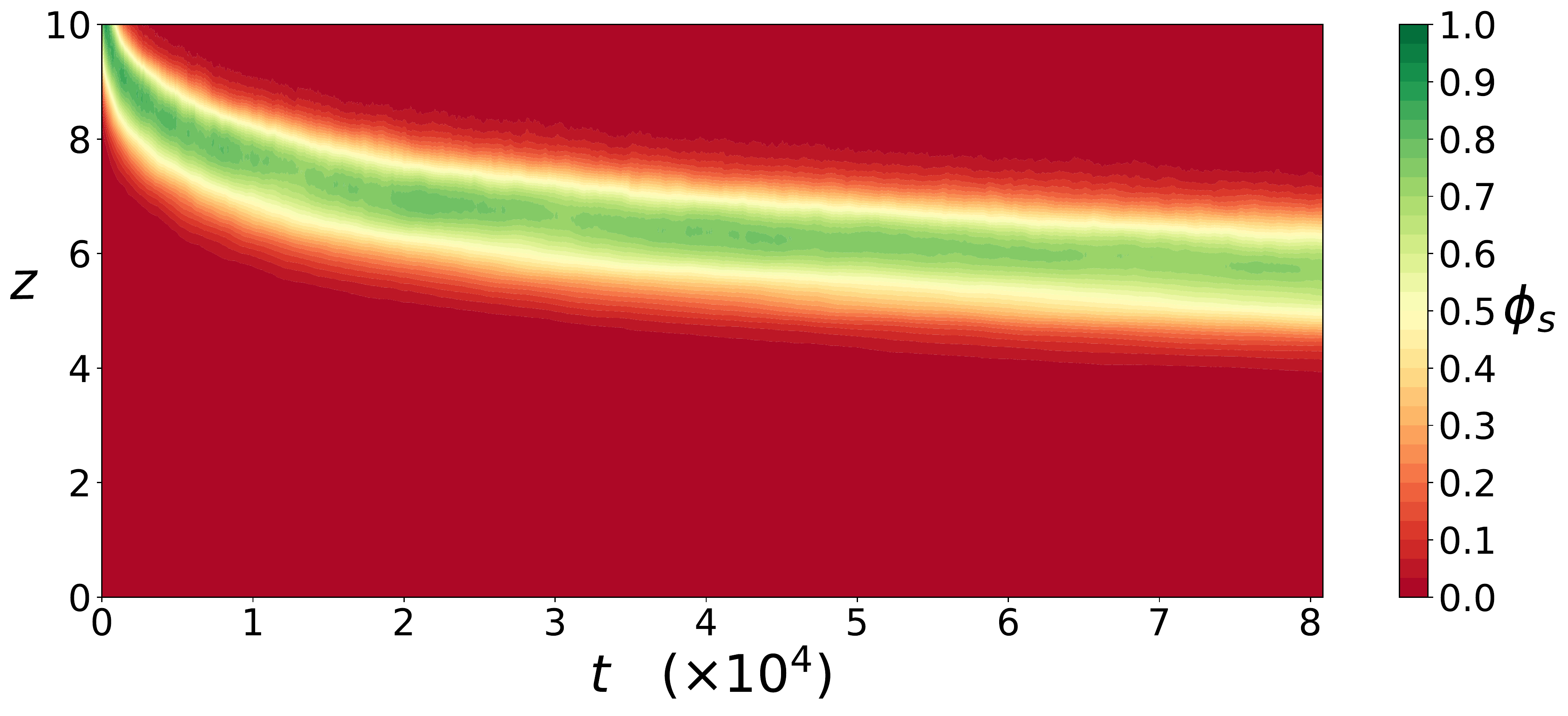}}\\
 \subfloat[]{\includegraphics[height=5cm]{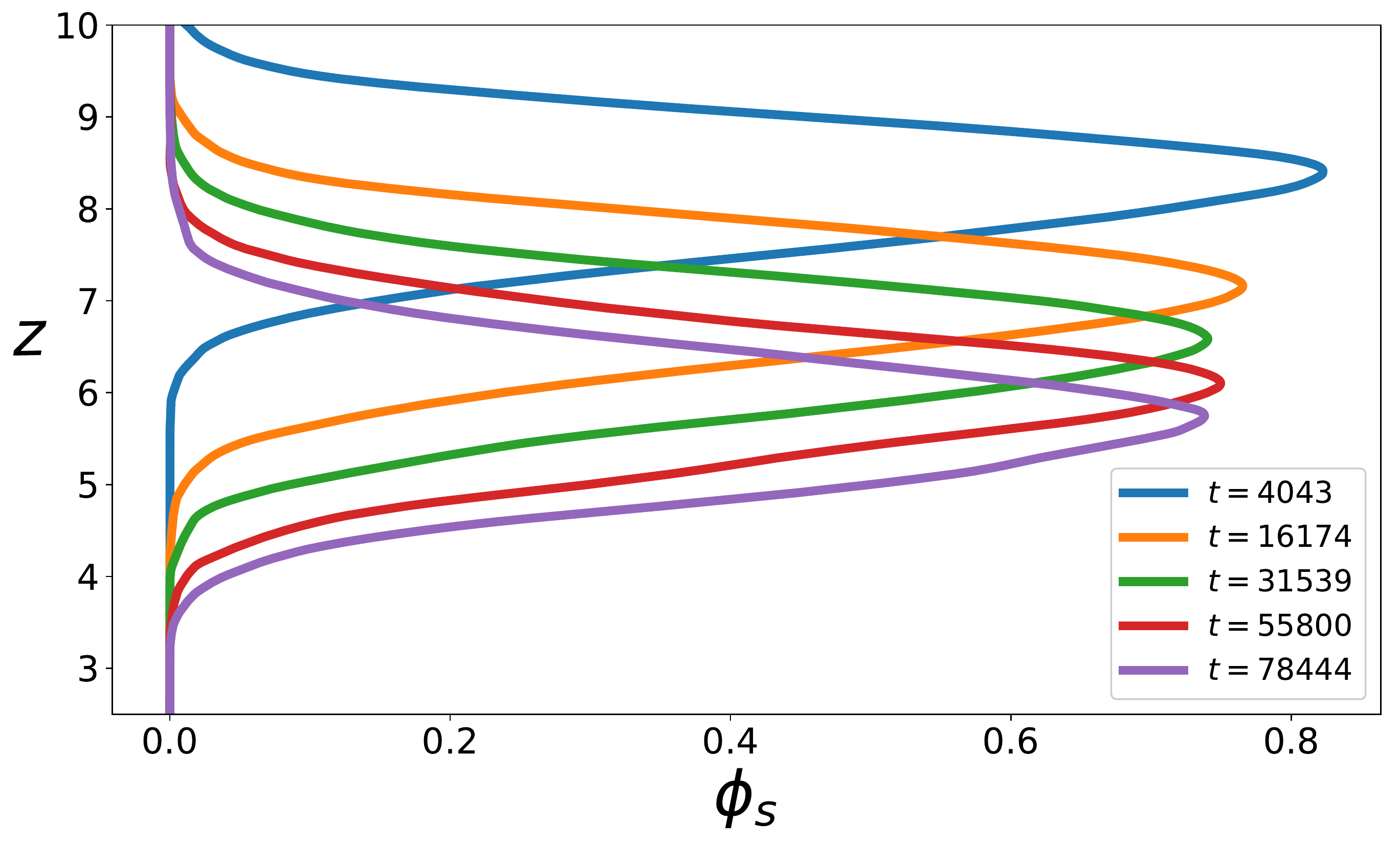}}
 \caption{(a) Numerical setup, (b) Typical temporal evolution of fine particles concentration profile $\phi_s$ for the case $N_s=2$, (c) Concentration profiles in small particles for different times. 
}
 \label{fig:num_setup}
\end{figure}

The numerical setup is presented in figure~\ref{fig:num_setup}a. In the following, subscripts $l$ and $s$ denote quantities for large and small particles respectively. Initially, large particles of diameter $d_l = 6$ mm and fine particles of diameter $d_s=4$ mm (size ratio $r=1.5$) are deposited by gravity over a rough fixed bed made of large particles. The size of the 3D domain is $30d_l \times 30 d_l$ in the horizontal plane in order to have converged average values \citep{maurin2015} and is periodic in the streamwise and spanwise directions. The number of particles of each class is assimilated to a number of layers, $N_l$ and $N_s$. It represents in term of particle diameters the height that would be occupied by the particles if the packing fraction was exactly 0.61. 
The height of the bed at rest is thus defined by $H = N_ld_l+N_sd_s$ and $H$ is fixed to $10d_l$ in all the simulations. The number of layers of fine particles $N_s$ varies from $0.01$ (only a few small particles) up to $2$ layers (corresponding to figure~\ref{fig:num_setup}a), while $N_l$ changes accordingly in order to keep $H=10d_l$.
The bed slope is fixed to $10\%$ ($5.7^\circ$), representative of mountain streams. The Shields number is defined as the dimensionless fluid bed shear stress $\theta = \tau_f/[(\rho^p -\rho^f)gd_l]$ and simulations are performed at $\theta \simeq 0.1$. For this purpose, the free surface position is fixed and corresponds to a water depth of $h=3.1d_l$. 

At the begining of each simulation, the fluid flows by gravity and sets particles into motion. A first transient phase takes place, during which fluid and particles are accelerating. During this period, segregation is very fast and at the end of the transient phase, the small particles have already infiltrated into the first layers of the bed. In this paper, the study focuses on the dynamics of segregation once the system is at transport equilibrium and small particles have reached the quasi-static layer.

 The horizontal averaged volume fraction per unit granular volume of small (resp. large) particles is defined as $\phi_s$ (resp. $\phi_l$). By definition, the two volume fractions sum to unity,
\begin{equation}
 \phi_s+\phi_l= \dfrac{\Phi_s}{\Phi_s+\Phi_l} + \dfrac{\Phi_l}{\Phi_s+\Phi_l} = 1,
 \label{eq:sum_phi}
\end{equation}
where $\Phi_s$ (resp. $\Phi_l$) is the volume the volume fraction of small (resp. large) particles defined per unit mixture volume as in the previous section. Lastly,  $z_c$ is defined as the vertical position of the center of mass of small particles and $dz_c/dt$ as its velocity.

The results are presented without dimension and the following dimensionless variables are considered,
\begin{equation}
 \begin{array}{ccc}
 \tilde{z} = \dfrac{z}{d_l}, & \tilde{t} = t\sqrt{g/d_l}, & \widetilde{\left<v_x\right>^p} = \dfrac{\left<v_x\right>^p}{\sqrt{gd_l}}.
 \end{array}
 \label{eq:adim}
\end{equation}
In the following, the tildes are dropped for sake of clarity.

\subsection{Results}

Simulations have been performed for different numbers of layers of small particles $N_s$. Figure~\ref{fig:num_setup}b shows the temporal evolution of fine particle concentration profiles for the case $N_s=2$, corresponding to two layers of small particles deposited on top of a layer made of large particles. At the beginning, the small particles infiltrate rapidly into the first few layers of large particles. As small particles infiltrate downward, large particles rise to the surface. The DEM simulations exhibit a two layer structure, with small particles sandwiched between two layers of large particles. While infiltrating, the thickness of the small particle layer gets slightly larger. Profiles of concentration for different times are presented on figure~\ref{fig:num_setup}c. The concentration profiles exhibit a Gaussian-like shape. After the transient phase, neither the maximal value nor the width of the profiles evolve in time, suggesting that the small particles infiltrate the bed as a layer having a constant thickness and that this layer is just convected downward by segregation inside the layer of large particles. 

\begin{figure}
\centering
 \subfloat[]{\includegraphics[height=6cm]{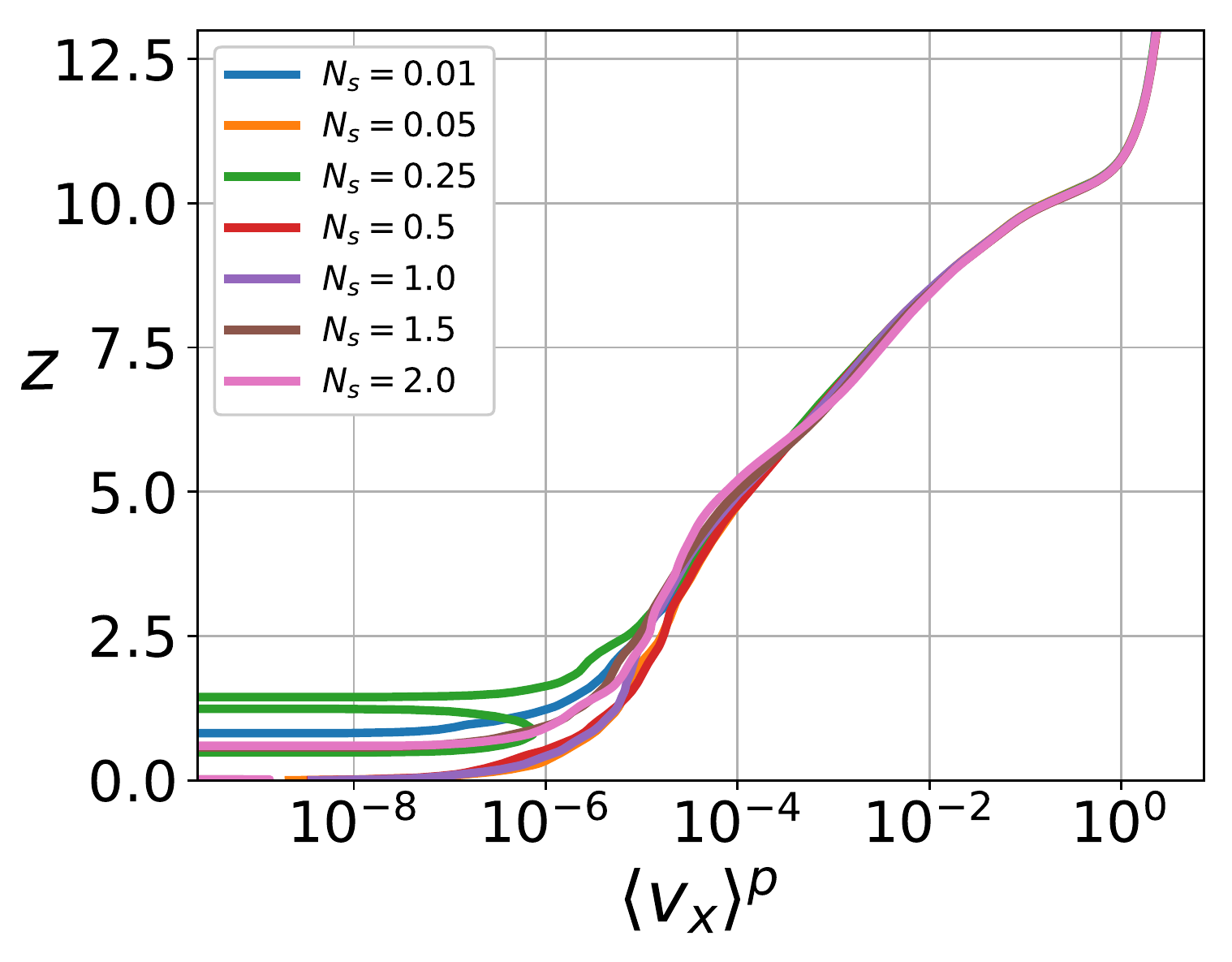}}\\
 \subfloat[]{\includegraphics[height=6cm]{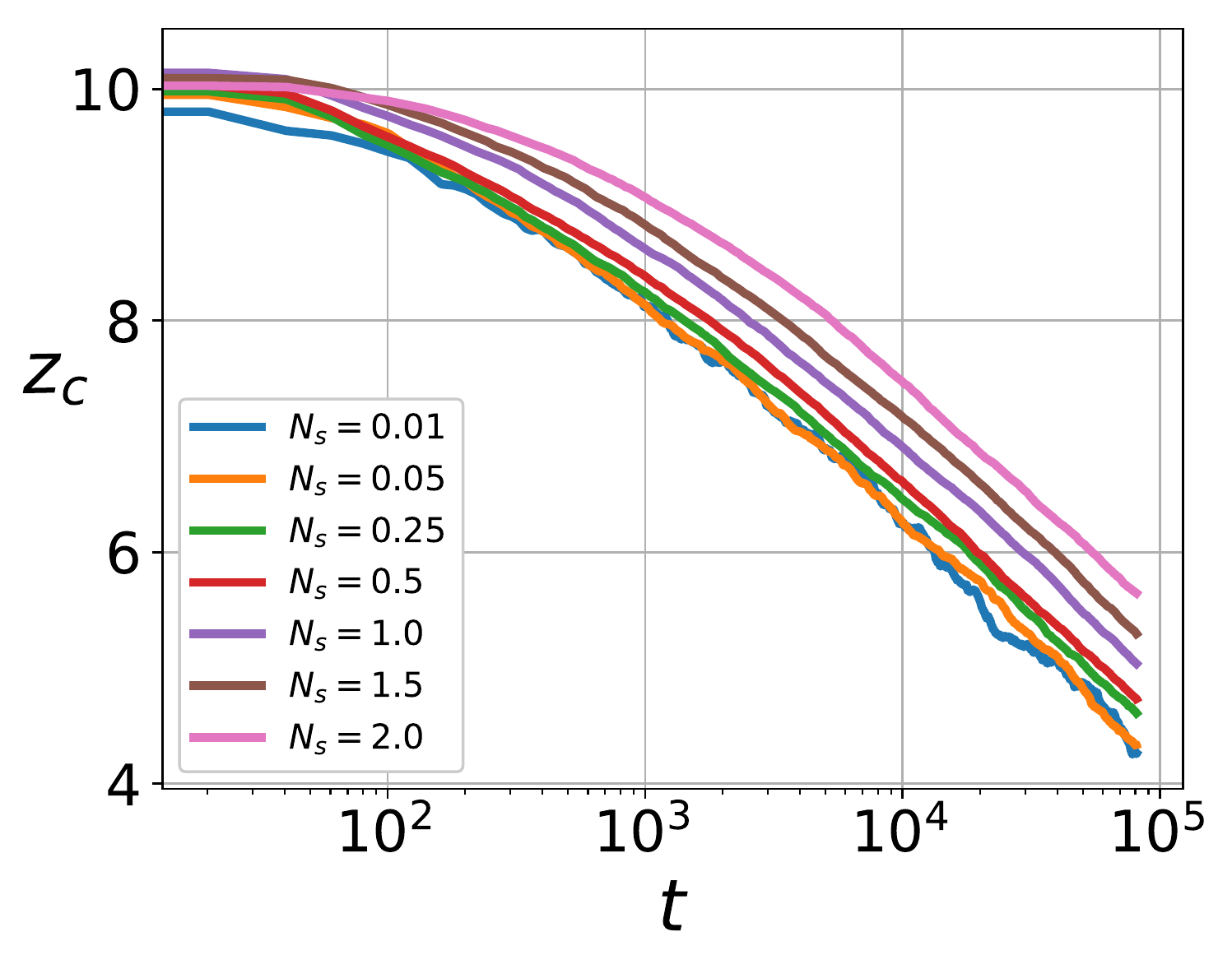}}
 \caption{(a) Streamwise space time averaged particle velocity as a function of the height, (b) evolution of the vertical position of the mass center of small particles with time}
 \label{fig:results}
\end{figure}

The vertical time and volume-averaged streamwise velocity profile of the particle mixture is plotted in figure~\ref{fig:results}a for all the simulations. All the curves are superimposed meaning that the response of the granular medium to the fluid flow forcing is not modified by the number of small particles. The granular forcing is therefore the same for all the simulations independently of the number of small particles in the initial condition. In the quasi-static region (i.e. between $z = 3$ and $8$ approximately), the linearity of the profiles in the semilog plot indicates that the velocity is exponentially decreasing in the bed. Due to the presence of a fixed layer of particles at the bottom of the domain, the velocity goes to zero there. It is interesting to note that the entire bed is in motion, even if the velocity can be very low at the bottom of the quasi-static region. This is characteristic of a creeping flow.

%

Since the small particles infiltrate the bed as a layer, the center of mass of the small particles, $z_c$, is a representative position of the entire layer. Figure~\ref{fig:results}b shows the temporal evolution of this position for all the simulations, in a semi-logarithmic plot. After the initial transient phase, the curves become linear, meaning that $z_c$ is a logarithmic function of time
\begin{equation}
 z_c = -a\ln(t)+b.
 \label{eq:fit_zc}
\end{equation}
In equation~\eqref{eq:fit_zc}, the coefficient $a$ corresponds to the absolute slope of the curve and characterises the segregation velocity ($dz_c/dt = -a/t$). Whatever the  number of small particles in the simulation, figure~\ref{fig:results}b shows that all the curves are parallel to one another, meaning that $a$ is independent of the number of layers of small particles, $N_s$. Therefore the segregation velocity $dz_c/dt$ is also independent of the number of layers of small particles.\\

In the following, these simulations will be analysed using the dimensional analysis presented in the introduction (equation~\eqref{eq:piTh3}) with the aim to confirm the dependence of the segregation flux on the inertial number $I$ and on the local concentration $\phi_s$.

\subsection{Dependence on the inertial number}

\begin{figure}
\centering
 \subfloat[]{\includegraphics[height=6cm]{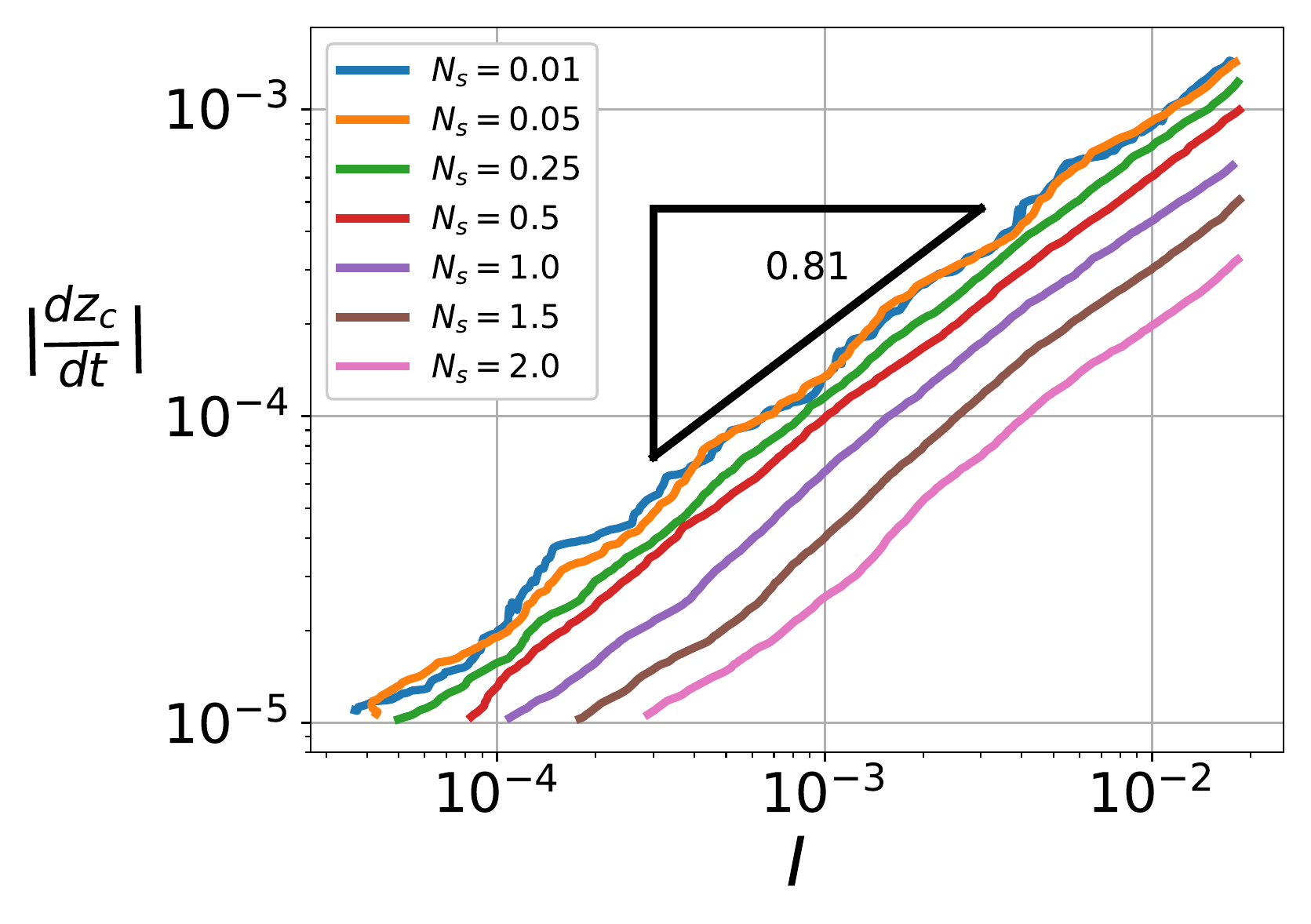}}\\
 \subfloat[]{\includegraphics[height=6cm]{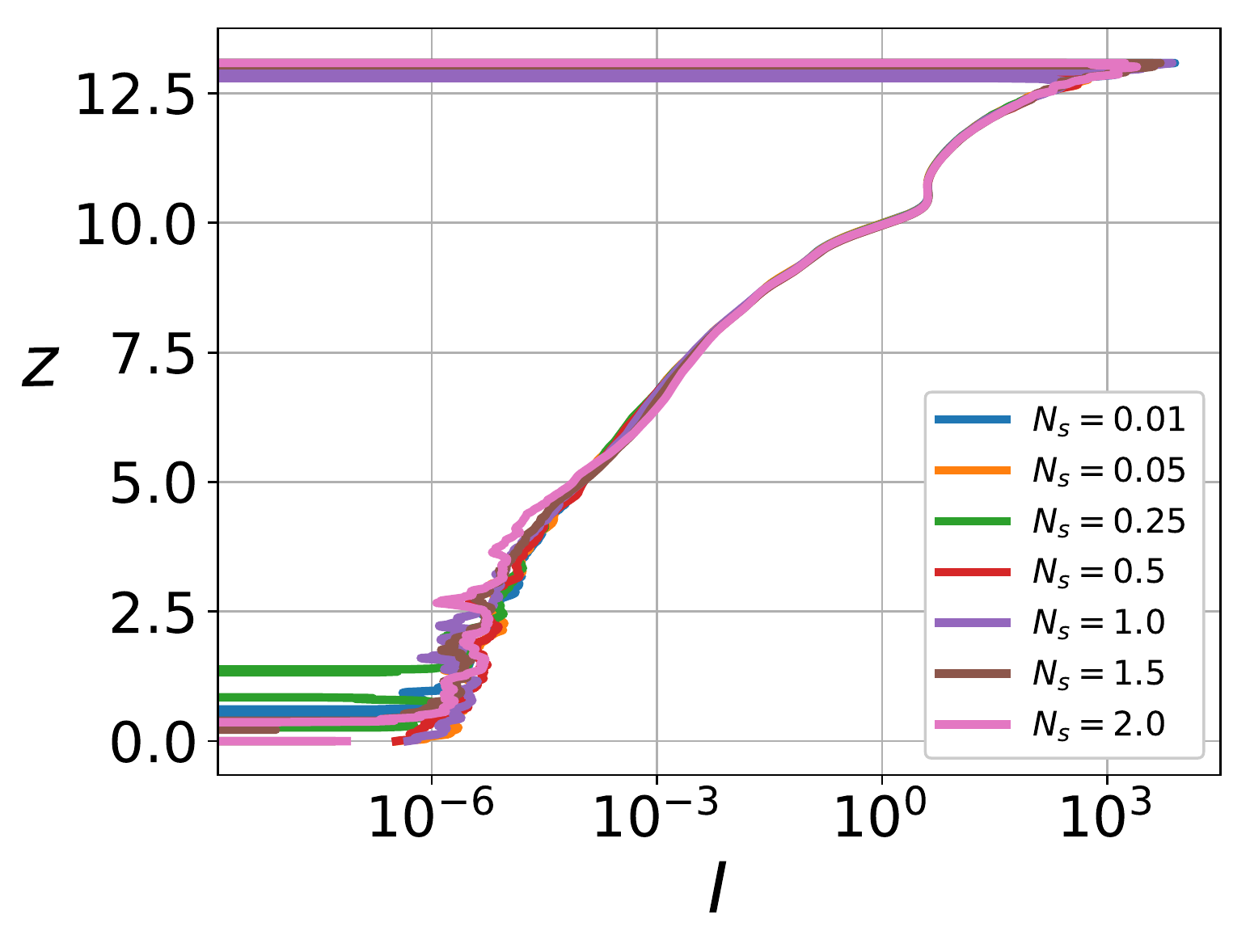}}
 \caption{(a) Segregation velocity dependence on the local inertial number, (b) inertial number profile}
 \label{fig:SegI}
\end{figure}

In figure~\ref{fig:SegI}a, the dimensionless segregation velocity is plotted against the large particle inertial number $I=\dot{\gamma}^pd_l/\sqrt{P/\rho^p}$. The segregation velocity is higher for larger inertial number. The linearity of the curves shows that the segregation velocity is indeed a power law of the inertial number. For all the simulations the same exponent is obtained and a best fit of the curves gives a value of $0.81$ for the exponent,
 
 \begin{equation}
  \dfrac{dz_c}{dt} \propto I^{0.81}.
  \label{eq:I_power}
 \end{equation}

Interestingly, \cite{fry2018} obtained a very similar result for dry granular flows at higher inertial numbers ($I \in [10^{-2}, 1 ]$). They obtained an exponent $0.84$, very close to the $0.81$ exponent obtained in the present configuration. Our results suggest that the behaviour observed by \cite{fry2018} is valid in a wider range of inertial numbers, in particular, in the quasi-static regime ($I \in [10^{-5}, 10^{-1} ]$).

Relation~\eqref{eq:I_power} allows understanding the temporal evolution of the fine particles center of mass observed previously. Figure~\ref{fig:SegI}b shows that the inertial number profile is an exponential function of $z$ as well as $I^{0.81}$ which can be written as 
\begin{equation}
I^{0.81} = I_0e^{z/c}.
\label{eq:Ifit}
\end{equation}
Introducing~\eqref{eq:Ifit} into expression~\eqref{eq:I_power} leads to
\begin{equation}
 \dfrac{dz_c}{dt} =-b_1 e^{z_c/c},
 \label{eq:SegI}
\end{equation}
with $b_1$ a positive constant. Equation~\eqref{eq:SegI} can be integrated
\begin{equation}
  \int e^{-z_c/c}dz_c = \int -b_1 dt,
\end{equation}
leading to
\begin{equation}
  ce^{-z_c/c} = b_1t + Const \sim b_1 t,
\end{equation}
for long times. It can be rewritten as
\begin{equation}
  z_c = -c\ln(t) + b_2,
  \label{eq:SL2}
\end{equation}
where $b_2 = -c\ln(b_1/c)$ is a constant. This analysis shows that the logarithmic descent of the small particle center of mass is a consequence of the dependence of the segregation velocity on the inertial number. This is confirmed by the comparison between the coefficients $a$ in equation~\eqref{eq:fit_zc} and $c$ in equation~\eqref{eq:Ifit}, that should be equal. Fitting of the elevation of the center of mass (figure~\ref{fig:results}b) and of $I^{0.81}$ yields coefficients $a$ and $c$ (see table~\ref{table:SLcoef}). The maximal difference between $a$ and $c$ is around $3\%$ confirming the present analysis and showing that the inertial number is indeed the controlling parameter. 

\begin{table}
\begin{center}
\def~{\hphantom{0}}
\begin{tabular}{lcccc}
		
        $N_s$   & $a $  & $c$ & error (\%) \\
		$0.01$  & $0.863$ & $0.873$   & $1.159$\\
		$0.05$  & $0.850$ & $0.864$   & $1.647$ \\
		$0.25$  & $0.823$ & $0.847$   & $2.916$ \\
		$0.5$   & $0.843$ & $0.862$   & $2.254$ \\
		$1$     & $0.833$ & $0.857$   & $2.800$ \\ 
		$1.5$   & $0.820$ & $0.828$   & $0.976$ \\
		$2$     & $0.849$ & $0.828$   & $2.473$ \\
\end{tabular}
\caption{Values of coefficients $a$ (slope of the center of mass) and  $c$ (exponential decay of the inertial number to the power $0.81$) obtained from fitting the curves of figures \ref{fig:results}b and \ref{fig:SegI}b and the error in percentage.}
\label{table:SLcoef}
\end{center}
\end{table}





\subsection{Bottom controlled segregation}\label{sec:bottom}

While this analysis clearly explains the trends observed, figure~\ref{fig:SegI}a shows that for a given value of the inertial number, different segregation velocities are obtained depending on the initial number of small particles, $N_s$. 

\begin{figure}
 \centering
 \subfloat[]{\includegraphics[height=6cm]{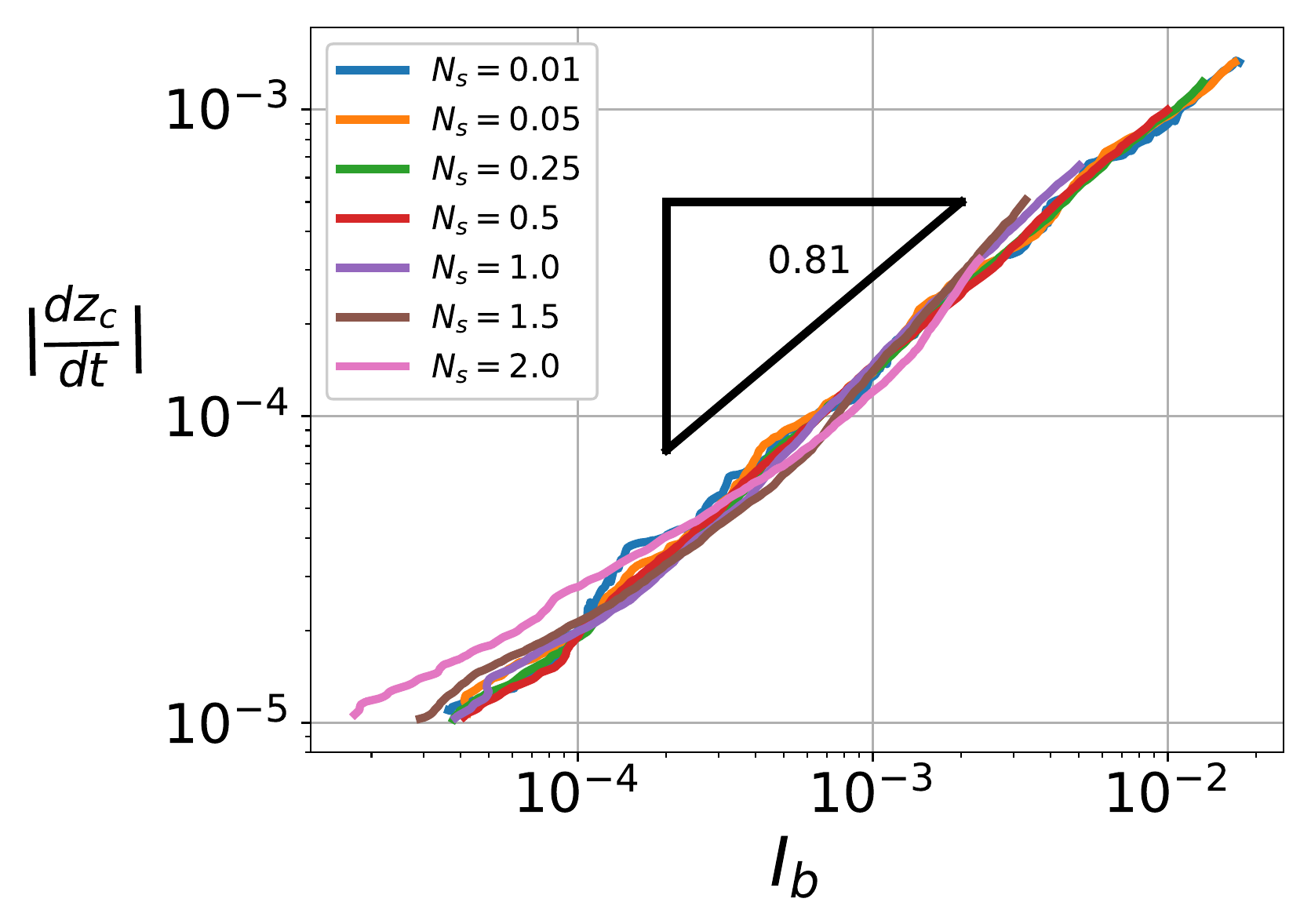}}\\
 \subfloat[]{\includegraphics[height=6cm]{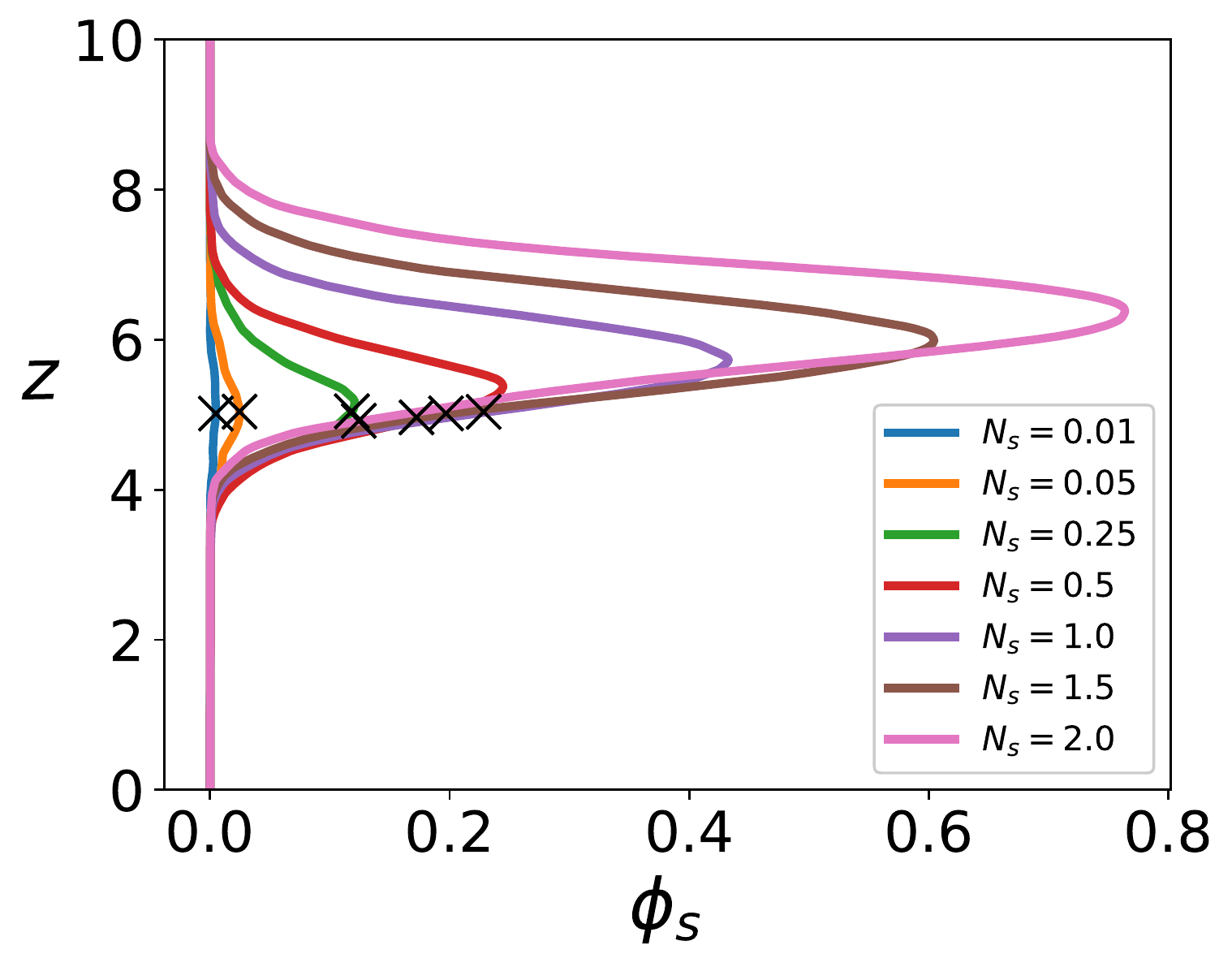}}
 \caption{(a) Center of mass velocity dependence on the inertial number at the bottom of the layer of small particles. (b) Small particles concentration profiles at time $t=40435$ and vertical position of the bottom of the layer ($\times$).}
 \label{fig:SegIBot}
\end{figure}

Since the inertial number follows an exponential profile, it varies importantly throughout the layer of small particles. Thus according to equation~\eqref{eq:I_power} the segregation velocity should be lower at the bottom than at the top of the layer. This lower segregation velocity at the bottom of the layer implies that all the small particles above cannot move downward faster than the lowest particles in the layer. Defining the position of the bottom of the layer as $z_{b} = z_c-W/2$, with $W = 2N_sd_s/d_l$ the small particle layer thickness, figure~\ref{fig:SegIBot}a shows the dependence of the segregation velocity on the inertial number at the bottom of the layer. All the curves collapse on a master curve and a power law relationship is found between the segregation velocity and the inertial number at the bottom of the layer $I_b =I(z_{b})$, 
\begin{equation}
 \dfrac{dz_c}{dt} = -\alpha_0 I_{b}^{0.81},
 \label{eq:SLnew2}
\end{equation}
where $\alpha_0$ is a positive constant independent of the number of layers of small particles. This result shows that the segregation velocity of the layer is indeed completely controlled by the inertial number at the bottom of the layer and it does not contain any dependence on the number of small particle layers, $N_s$.\\

The layer thickness has been chosen to be two times the thickness it would occupy if only small particles were present, $W = 2N_sd_s/d_l$. This choice is motivated by the fact that the layer is formed of a mixture of both large and small particles. Figure~\ref{fig:SegIBot}a shows that for the different cases, the width obtained is indeed consistent with the actual thickness. Note that when $N_s$ tends to zero, $z_{b}$ tends to $z_c$ the center of mass of the small particles. In the extreme case where only one small particle is present, this position corresponds to its center. Figure~\ref{fig:SegIBot}b shows, at time $t=40435$, the profiles of small particle concentration, where the black crosses denote the position of the bottom of the layer. The lower limb of the concentration profiles are superimposed and the position of the bottom of the layer is identical for all tested values of $N_s$. Whatever the number of small particles, they pile up above the bottom position, where the segregation dynamics is controlled.\\

This analysis highlights the dependence of segregation on the inertial number in bedload transport. Furthermore, the bottom of the small particles layer is shown to be a key position that controls segregation. In particular, it explains why the small particles infiltrate the bed as a layer of constant thickness.

\subsection{Size ratio influence}\label{sec:ratio}

The analysis presented above provides a reference position that describes the segregation dynamics. In the following, the effect of the size ratio is investigated. The previous simulations were all performed at the same size ratio, $r=1.5$. According to the dimensional analysis, the following relationship is expected,
\begin{equation}
 \dfrac{dz_c}{dt} = -\alpha_0 (r-1)^l I_{b}^{0.81}.
 \label{eq:SLnew2}
\end{equation}

\begin{figure}
 \centering
 \subfloat[]{\includegraphics[height=6cm]{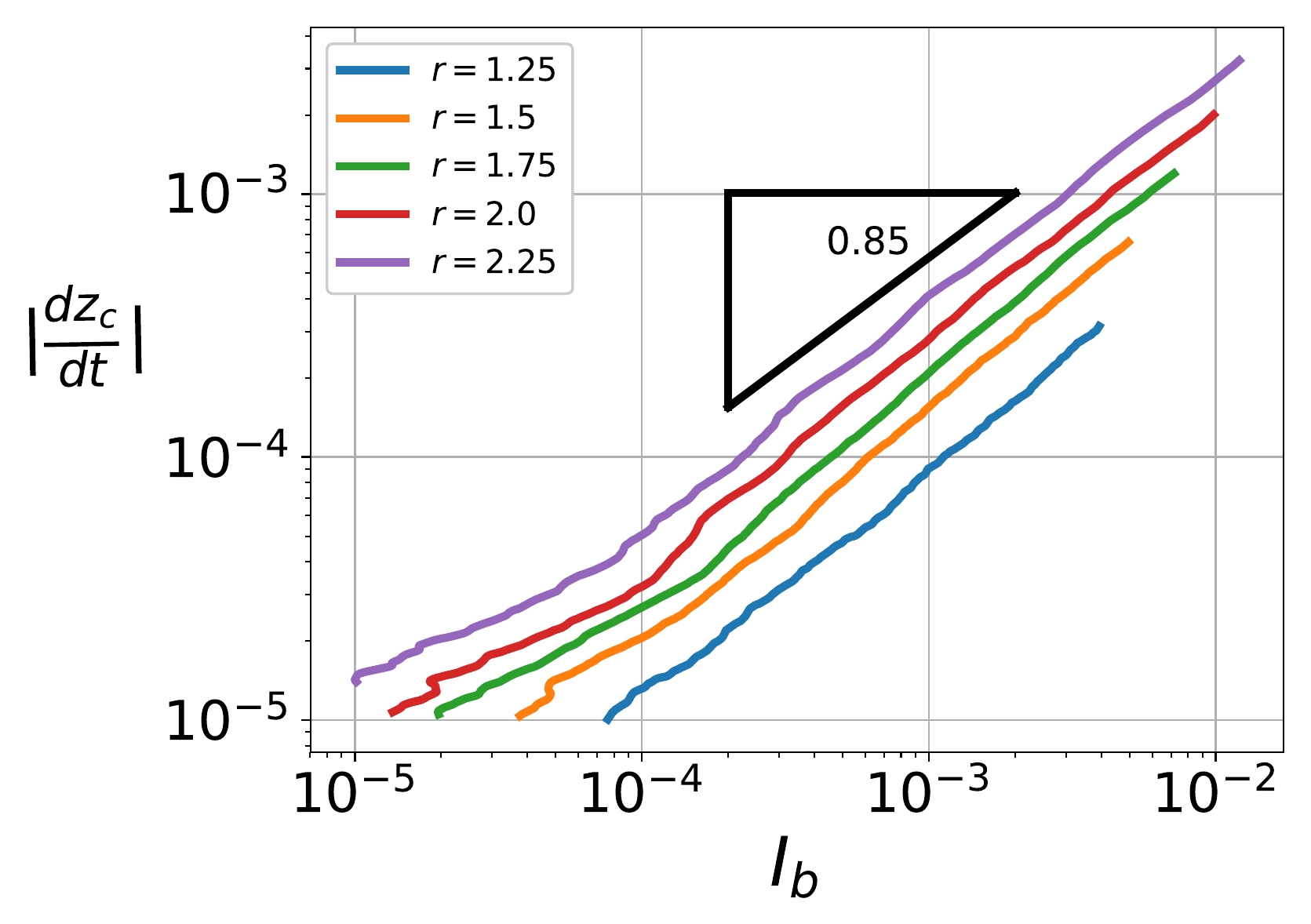}}\\
 \subfloat[]{\includegraphics[height=6cm]{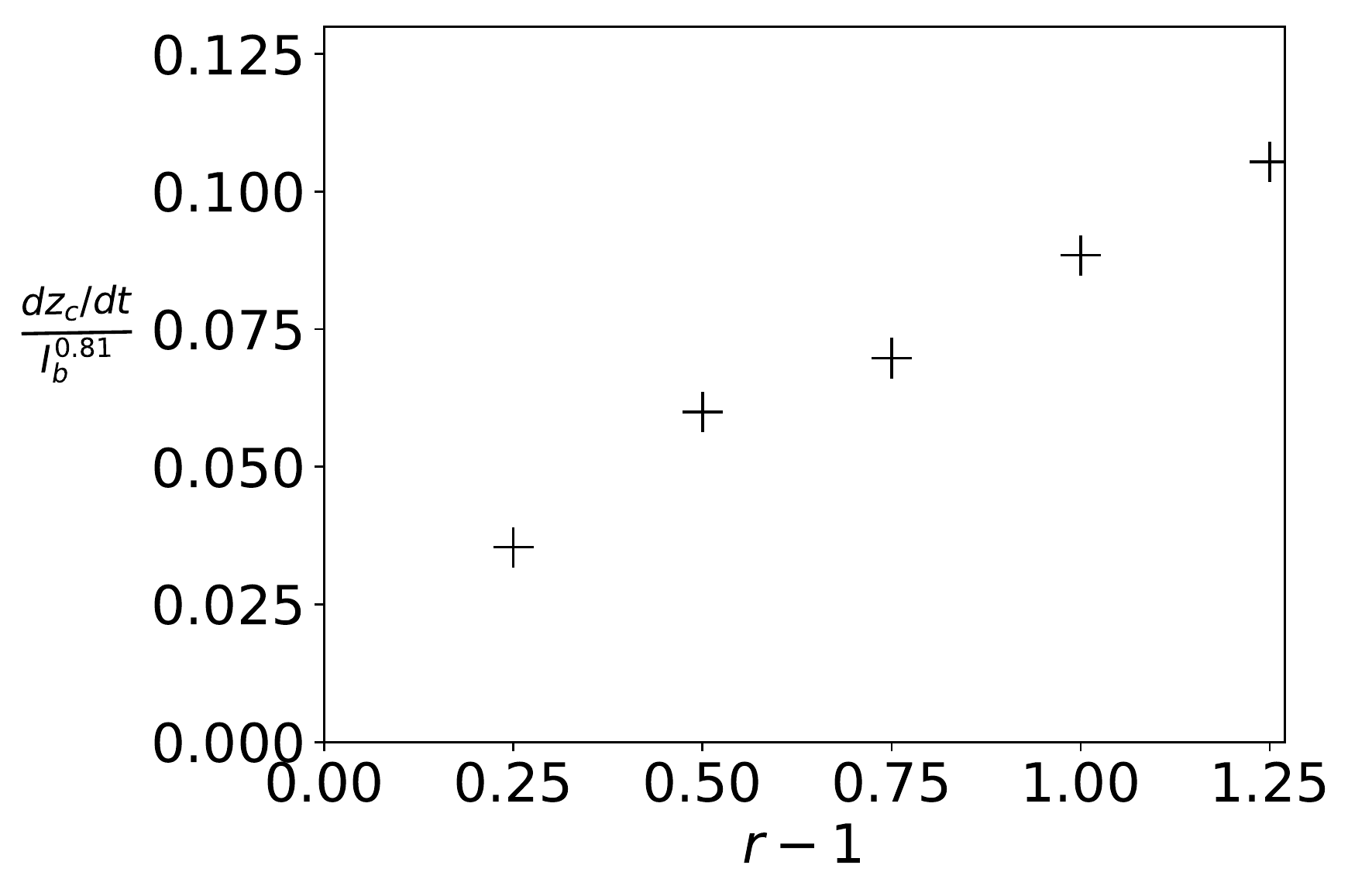}}
 \caption{(a): Center of mass velocity dependence on the inertial number at the bottom of the layer of small particles for different size ratio, (b): value of $\alpha_0$ with the size ratio.}
 \label{fig:size_ratio}
\end{figure}

A set of simulations in which the diameter of the small particles is varied was performed with a constant number of small particles layer $N_s=1$. It covers a range of size ratio from $r=1.25$ to $2.25$. The dependence of the center of mass velocity on the inertial number at the bottom of the layer of small particles is plotted on figure~\ref{fig:size_ratio}a. A similar power law relationship is recovered (the exponent is slightly larger than previously), showing that the inertial number is still the controlling parameter. The curves are parallel but not superimposed, inducing that, as expected, the size ratio plays an important role in the segregation dynamics. This dependence is presented in figure~\ref{fig:size_ratio}b where the ratio between the segregation velocity and the inertial number to the power $0.81$ is plotted as a function of the size ratio minus one. The dimensionless segregation velocity normalised by $I_b^{0.81}$ evolves linearly with $r-1$ suggesting that the exponent $l$ in~\eqref{eq:SLnew2} equals to unity. However the maximal efficiency at $r=2$ reported by \cite{golick2009}, \cite{thornton2012} and \cite{guillard2016} is not recovered in this configuration as the segregation velocity still increases for $r > 2$. This difference with the literature may be due to the fact that segregation is studied here in the quasi-static part of the bed.
 
To summarize, the results results presented herein suggest that the segregation dynamics in bedload transport can be described, at first order, by the following relationship,
\begin{equation}
 \dfrac{dz_c}{dt} = -\alpha_0 (r-1) I_{b}^{0.81}.
 \label{eq:SLnew3}
\end{equation}

\subsection{Dependence on the local concentration $\phi_s$}\label{sec:concentration}


It is remarkable that the trends observed for the evolution of the center of mass and for the segregation velocity are independent of the number of small particle layers, $N_s$. Indeed, as the latter is increased from 0.01 to 2, the configuration moves from few isolated (non-interacting) small particles to two consistent layers of small particles infiltrating collectively inside the coarse bed. Therefore, it is expected that the increase of $N_s$, would increase the local particle concentration and reduce the segregation velocity due to particle hindrance effects.

This independence of the trend observed in the segregation velocity seems in contradiction with the literature \citep[e.g.][]{savage1988, dolgunin1995, vandervaart2015, jones2018}, in which a major influence of the local small particle concentration on the segregation velocity has been evidenced. In order to explain this apparent contradiction and get more insight into the physical processes at play, the results are discussed in the following as a function of local mechanisms within the framework of the continuous model of \cite{gray2005}, \cite{thornton2006} and \cite{gray2006}.

\section{Discussion in the framework of continuous modelling}\label{sec:mechanism}

\subsection{Local mechanism interpretation}\label{sec:advection}

In order to explain the apparent independence of the results on the concentration of small particles, a local analysis in the theoretical framework of continuous modelling is considered. The model of \cite{thornton2006} \eqref{eq:gray_init}, presented in the introduction, will be used as a baseline. In the present configuration, the velocity field has only a streamwise component (along $x$) and all quantities only depend on $z$. As a first step, diffusion is not considered and equation~\eqref{eq:gray_init} simplifies to a purely hyperbolic equation,
\begin{equation}
\dfrac{\partial \phi_s}{\partial t} - \dfrac{\partial F_s}{\partial z} = 0.
\label{eq:segmod}
\end{equation}

\begin{figure}
 \centering
 \centering
 \subfloat[]{\includegraphics[height=6cm]{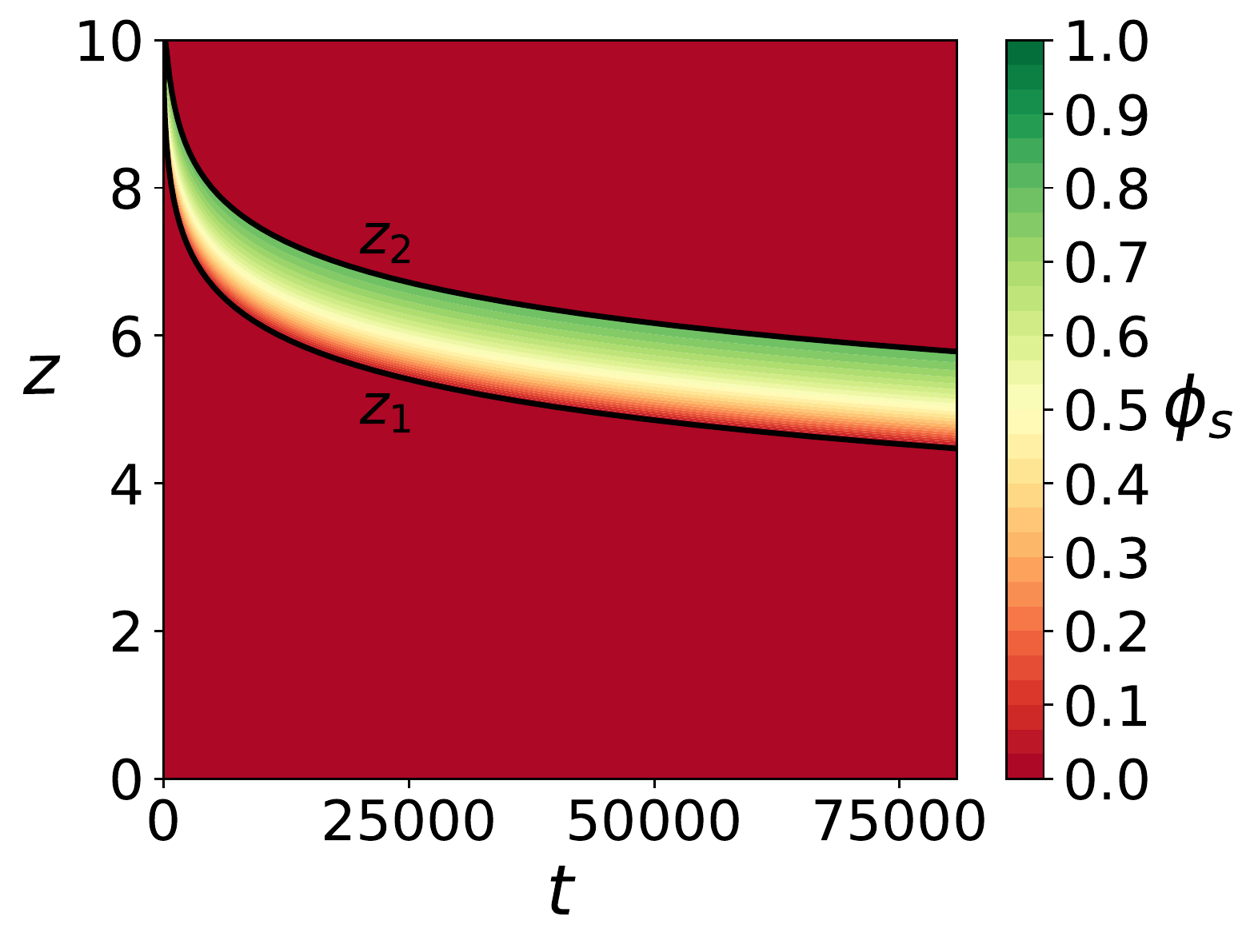}}\\
 \subfloat[]{\includegraphics[height=6cm]{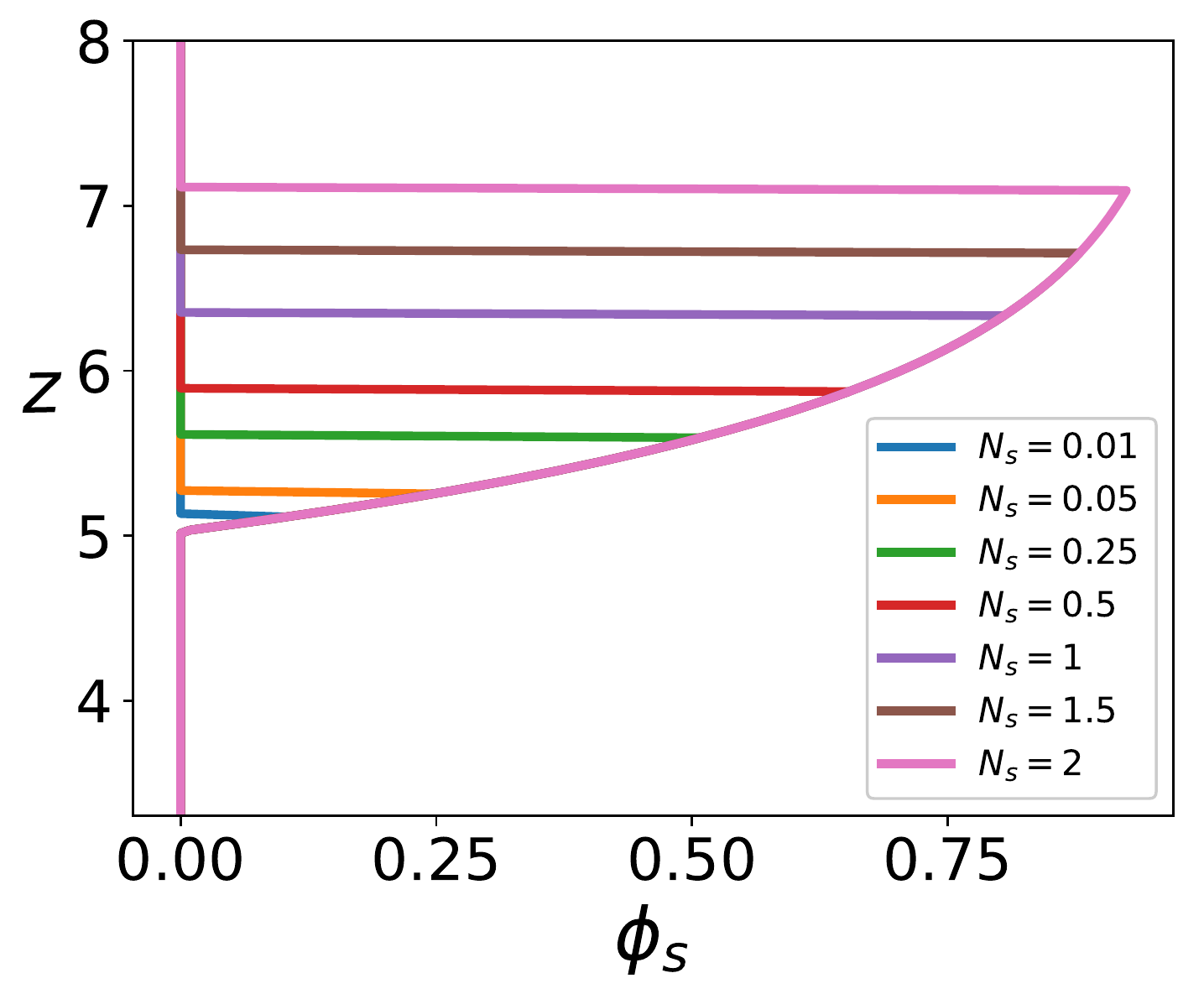}}
  \caption{(a) Case $N_s=1$, $r=1.5$ space time evolution of the concentration in fine particles $\phi_s$, (b) Profiles of concentration at time $t=80870$ with the continuum model without diffusion for several values of $N_s$. The same value of $S_r$ and $c$ have been used.}
 \label{fig:sol_adv}
 \end{figure}

Combining the form proposed in the introduction, the analysis of the DEM simulations~\eqref{eq:SLnew3} and the inertial number functional form~\eqref{eq:Ifit}, the segregation flux can be written as
\begin{equation}
 F_s = \alpha_0(r-1)I^{0.81}\phi_s(1-\phi_s) = S_{r0}e^{z/c}\phi_s(1-\phi_s),
 \label{eq:segflux}
\end{equation}
with $S_{r0} = \alpha_0(r-1)I_0$. The inertial number does not depend on the number of small particles (figure~\ref{fig:SegI}b) so $S_{r0}$ is independent on the initial number of small particles and only depends on the response of the granular mixture to the fluid flow forcing through $I_0$ and $c$. For $r=1.5$, the corresponding value of $S_{r0}$ obtained from the DEM simulations is: $S_{r0}=3.70\times 10^{-8}$. Due to the exponential decrease of the inertial number in the bed, the form of the segregation flux, $F_s$, is also exponential with $z$. \cite{may2010} already proposed an analytical solution for an exponential flux, but the flow was shear-driven from the bottom. Proceeding similarly, the analytical solution can be derived using the method of characteristics (appendix~\ref{appA}) for which the initial condition is simply a step of concentration representing an initial pure layer of small particles lying on a bed made of large particles.
\begin{equation}
 \phi_s(z, t=0) =
 \left\{
 \begin{array}{lr}
  0, &  \quad z < z_i,\\
  1, &  \quad z \geq z_i,
 \end{array}
 \right.
 \label{eq:sol_init}
\end{equation}
 with $z_i = N_l d_l$. The solution is plotted in figure~\ref{fig:sol_adv}a for the case $N_s=1$ and $r=1.5$. The initial discontinuity leads to an expansion fan delimited by two characteristics (see appendix~\ref{appA}). The first, denoted $z_1$, is going down and meets the bottom of the domain at time $t_1$, which is not reached during the simulations. The second is going up and is denoted $z_2$. When it meets the top of the domain at time $t_2$ (i.e. when the first large particle reaches the top boundary, which happens very rapidly), a shock is created that travels downward. It can be shown (see Appendix~\ref{appA}) that the analytical solution for time $t_2 \leq t \leq t_1$ is
 \begin{equation}
\phi_s(z,t)=
\left\{
 \begin{array}{lr}
 0, & z < z_1,\\
\phi_{fan}(z,t), & z_1 \leq z \leq z_2,\\
0, & z > z_2,
 \end{array}
 \right.
 \label{eq:long_time_sol}
\end{equation}
where $z_1$ is the characteristic curve~\eqref{eq:z_1} delimiting the bottom of the rarefaction zone, $z_2$ is given by the shock equation~\eqref{eq:z_2} delimiting the top of the rarefaction zone and $\phi_{fan}$ is the solution in the expansion fan given by~\eqref{eq:phi_fan},
\begin{equation}
 z_1(t) = z_i - c\ln\left(1+\frac{S_{r0}}{c}e^{z_i/c}t\right),
 \label{eq:z_1}
\end{equation}
\begin{equation}
 \dfrac{dz_2}{dt} = -S_{r0}e^{z_2/c}\left(1-\phi_s(z_2,t)\right),
 \label{eq:z_2}
\end{equation}
\begin{equation}
 \phi_{fan}(z,t) = \dfrac{1}{2} - \dfrac{c}{S_{r0}t}e^{-z/c} + \dfrac{c}{S_{r0}t}e^{-(z+z_i)/(2c)}\sqrt{1+\left(\dfrac{S_{r0}t}{2c}\right)^2e^{(z+z_i)/c}}.
 \label{eq:phi_fan}
\end{equation}
 
 Despite the complexity of the solution~\eqref{eq:long_time_sol}, some simplifications can be made for long times smaller than $t_1$. Indeed for long times, the square root term in the second equation of system~\eqref{eq:phi_fan} simplifies to
\begin{equation}
 1+\left(\dfrac{S_{r0}t}{2c}\right)^2e^{(z+z_i)/c} \sim \left(\dfrac{S_{r0}t}{2c}\right)^2e^{(z+z_i)/c},
\end{equation}
and the volume fraction profile $\phi_s$ in the rarefaction fan simplifies as,
\begin{equation}
 \phi_{fan}(z,t) = 1 - \dfrac{c}{S_{r0}t}e^{-z/c}.
 \label{eq:rarefaction_simple}
\end{equation}

Some simplifications can also be done on the boundary curves of the rarefaction zone $z_1$ and $z_2$. Introducing the long time solution~\eqref{eq:rarefaction_simple} in the shock equation of $z_2$~\eqref{eq:z_2}, a much more simple equation of $z_2$ is obtained for long times
\begin{equation}
 \dfrac{dz_2}{dt} = -\dfrac{c}{t},
\end{equation}
for which the solution is
\begin{equation}
 z_2(t) = -c\ln\left(t\right) + \beta,
 \label{eq:asymp_z2}
\end{equation}
where $\beta$ is an integration constant. Concerning the characteristic curve $z_1$, the term in the logarithm in~\eqref{eq:z_1} simplifies for long times to
\begin{equation}
 1+\frac{S_{r0}}{c}e^{z_i/c}t \sim \frac{S_{r0}}{c}e^{z_i/c}t,
\end{equation}
and the following expression is obtained for $z_1$
\begin{equation}
 z_1(t) = -c\ln(t) -c\ln\left(\frac{S_{r0}}{c}\right).
 \label{eq:asymp_z1}
\end{equation}

Therefore, the upper and lower bounds of the small particle layer, $z_2$ and
$z_1$, both move down asymptotically as $-c\ln(t)$, and the layer thickness $z_2(t) - z_1(t)$ is constant. Lastly, the form of the flux considered in the continuum model $F_s \propto I^{0.81}\phi_s(1-\phi_s)$ implies that the vertical velocity of the small particles $w_s = -F_s/\phi_s$ is
 \begin{equation}
  w_s(z,t) = -\dfrac{S_{r0}}{I_0}I^{0.81}(z)(1-\phi_s(z,t)).
  \label{eq:ws_1}
 \end{equation} 
 Using equation~\eqref{eq:rarefaction_simple}, the vertical velocity can be expressed for long times as
 \begin{equation}
  w_s(z) = -\dfrac{S_{r0}}{I_0}I^{0.81}\dfrac{c}{S_{r0}t}e^{-z/c},
 \end{equation}
and remembering that $I^{0.81} = I_0e^{z/c}$,
\begin{equation}
  w_s(z) = -\dfrac{c}{t}.
  \label{eq:ws_2}
 \end{equation}
 
Considering long time solutions and an imposed forcing by the inertial number, the model exhibits the exact same behavior observed in the DEM simulations. First, the small particles move down as a layer into the coarse particles, with the lower and upper bound of the small particles layer showing a logarithmic decrease with time (\ref{eq:asymp_z2} and \ref{eq:asymp_z1}). This decrease at a common slope $-c$ related to the forcing, imposes a constant layer thickness with time, as observed in the DEM simulations. The consequence is that the segregation velocity of the small particles is the same throughout the layer~\eqref{eq:ws_2}. Considering the formulation of the downward velocity of small particles~\eqref{eq:ws_1}, this means that the product of the solid volume fraction and the inertial number is constant within the small particle layer,
\begin{equation}
  w_s(z,t) = -\dfrac{S_{r0}}{I_0}I^{0.81}(z)(1-\phi_s(z,t)) = -c/t.
  \label{eq:ws_3}
 \end{equation}
Therefore, at a given time, the shape of the small particle concentration profile responds directly to the variation of the inertial number as a function of $z$. At a point within the layer, the decrease of inertial number observed when going inside the bed is compensated by an increase of the term $1-\phi_s$, and so a decrease of the solid volume fraction. This competition between the effect of the inertial number and the local concentration results in the shape of solution profiles presented in figure~\ref{fig:sol_adv}b.\\

In order to corroborate the explanation of the mechanisms at play, the lower bound of the small particle layer $z_1$ for which $\phi_s = 0$ is considered. It corresponds, in figure~\ref{fig:sol_adv}b, to the position where all profiles collapse to zero at the bottom. At this position, the segregation velocity is only a function of the inertial number at the bottom of the layer:
\begin{equation}
w_s(z,t) = -\dfrac{S_{r0}}{I_0}I^{0.81}(z_1) = -c/t
\end{equation}
This corresponds exactly to the mechanisms evidenced in the DEM  simulations in \S\ref{sec:bottom}, where the segregation velocity observed in the simulations have been shown to collapse as a function of the inertial number at a position considered to be the bottom of the small particle layer. The correspondance between the inertial number at the bottom position $z_b$ observed in DEM and the one calculated from the analytical continuum model $z_1$ (table~\ref{table:bot_comp}) confirms that the mechanisms evidenced here are indeed the one at play in the DEM simulations.

\begin{table}
\begin{center}
\def~{\hphantom{0}}
\begin{tabular}{cccccccc}
        $N_s$   & $0.01$  & $0.05$ & $0.25$  & $0.5$ & $1$ & $1.5$ & $2$\\
		$\dfrac{|z_1 -z_b|}{z_1}(\%)$  & $0.28$ & $0.22$  & $0.50$ & $0.09$ & $0.04$  & $0.73$ & $0.95$\\
\end{tabular}
\caption{Mean error between the bottom position from the DEM simulations $z_b$ and from the continuum model $z_1$. The maximal error of less than $1\%$ between both positions suggests that they correspond.}
\label{table:bot_comp}
\end{center}
\end{table}

Finally it is interesting to remark on figure~\ref{fig:sol_adv}b, that all the profiles are going to zero at the exact same position meaning that the bottom position is the same whatever the quantity of small particles. This confirms the DEM result that the segregation velocity is independent of $N_s$. Indeed, if there are more small particles, they pile up above the others without changing the position of the bottom where the segregation dynamics is controlled.

\begin{figure}
 \centering
 \includegraphics[height=6cm]{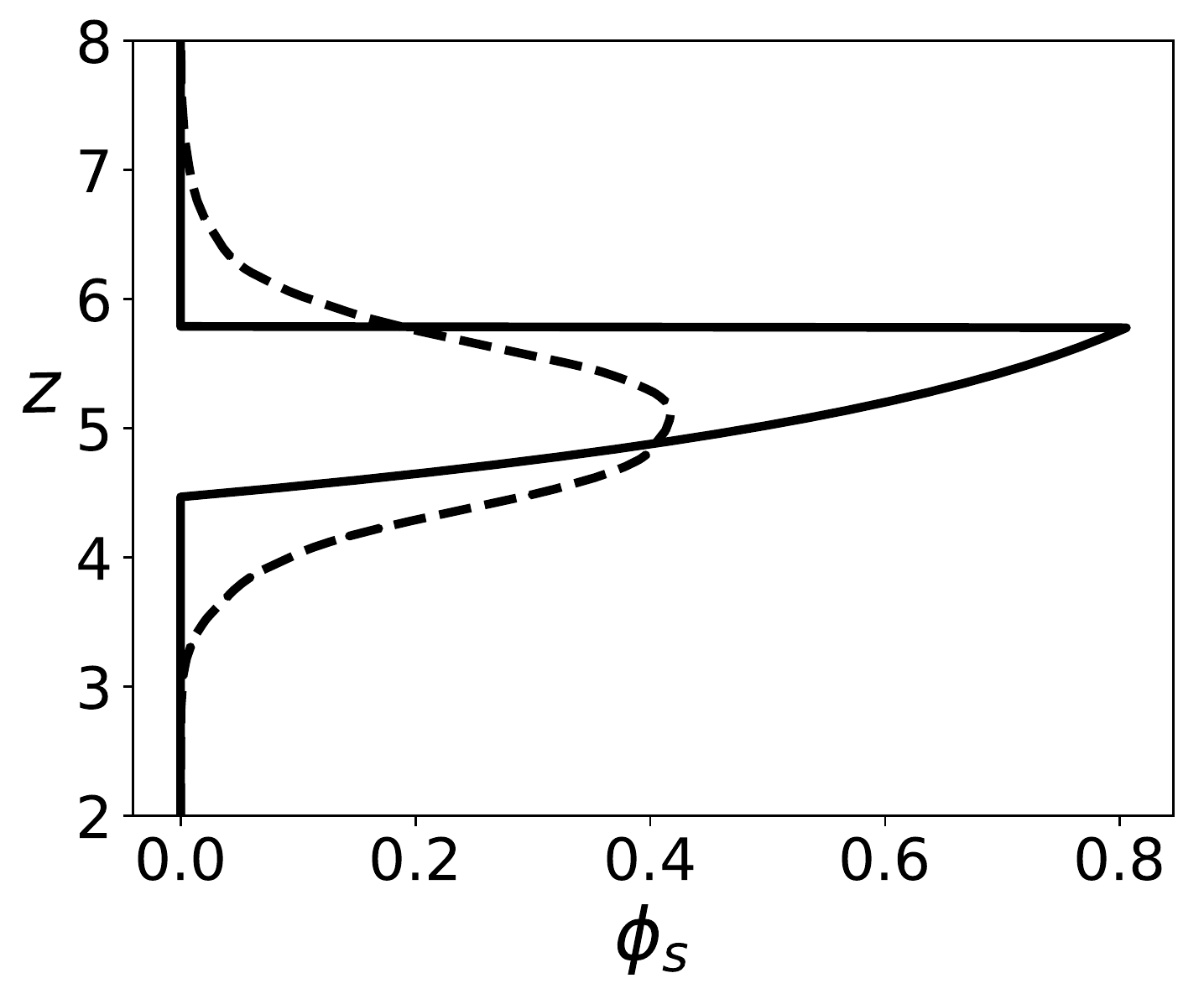}
 \caption{Comparison between DEM simulations (\protect\tikz[baseline]{\protect\draw[line width=0.3mm,densely dashed] (0,.5ex)--++(0.65,0) ;}) and segregation model (\protect\tikz[baseline]{\protect\draw[line width=0.3mm] (0,.5ex)--++(0.65,0) ;}) of the profile of concentration $\phi_s$ at time $t=80870$.}
 \label{fig:comp_profile}
\end{figure}

The continuum model without diffusion enabled understanding of the local segregation mechanisms and gave insights into the underlying physical mechanisms. While this model gives a good agreement at first order, as shown in figure~\ref{fig:comp_profile}, it clearly lacks diffusion in order to reproduce quantitatively the DEM simulations.

\subsection{Diffusion and upscaling in a continuous framework}\label{sec:diffusion}

The continuum model used in the previous subsection can be expanded to account for diffusion \citep{gray2006}, in order to reproduce more quantitatively the DEM results. The equation to be solved is given by,
\begin{equation}
\dfrac{\partial \phi_s}{\partial t} - \dfrac{\partial F_s}{\partial z} = \dfrac{\partial}{\partial z} \left(D\dfrac{\partial \phi_s}{\partial z}\right),
\label{eq:segmod_diff}
\end{equation}
with a segregation flux of the form $F_s = S_{r0}e^{z/c}\phi_s(1-\phi_s)$ and a diffusion coefficient $D(z)$ which is supposed to depend only on the vertical coordinate $z$. In this paper, a long time asymptotical solution is presented. At long times, DEM simulations have shown that the shape of the concentration profile $\phi_s$ is self similar and is only advected downward like a travelling wave. Indeed, the layer of small particles moves down as a layer of constant thickness as $-c\ln(t)$ and the shape of the concentration profile does not evolve with time. The following change of variable is proposed to place the problem in the moving frame of the small particles,
\begin{equation}
 \begin{array}{lr}
        \tau = t, & \xi = z+c\ln(t).
     \end{array}
\end{equation}
Equation~\eqref{eq:segmod_diff} in the moving frame becomes
\begin{equation}
 \dfrac{\partial \phi_s}{\partial \tau} + \dfrac{c}{\tau}\dfrac{\partial \phi_s}{\partial \xi} - \dfrac{\partial}{\partial \xi}\left(\dfrac{S_{r0}}{\tau}e^{\xi/c}\phi_s(1-\phi_s)\right)
 = \dfrac{\partial}{\partial \xi}\left(D(\xi-c\ln(\tau))\dfrac{\partial \phi_s}{\partial \xi} \right).
 \label{eq:trwa}
\end{equation}

\begin{figure}
 \centering
 \subfloat[]{\includegraphics[width=.6\linewidth]{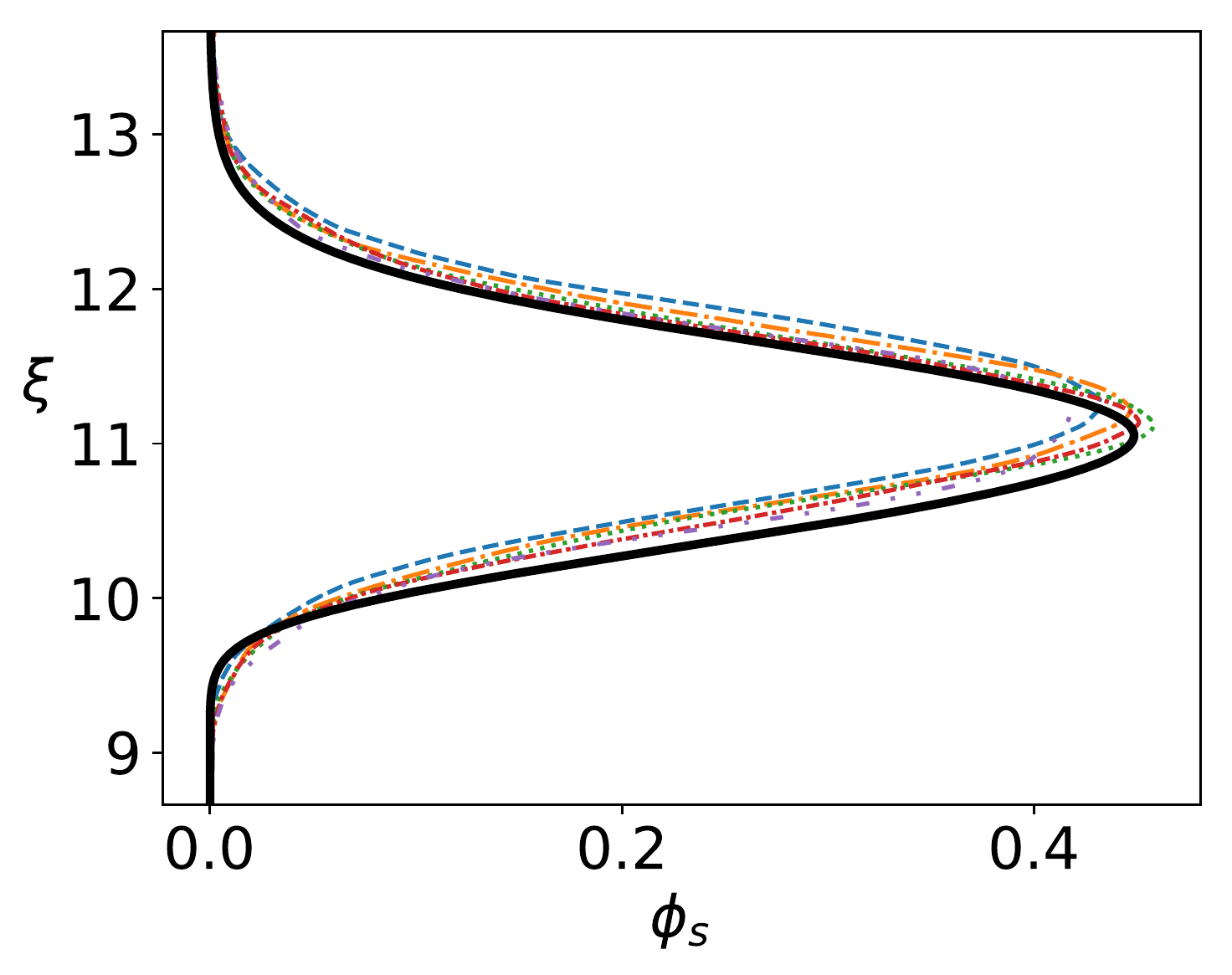}}\\
 \subfloat[]{\includegraphics[height=6cm]{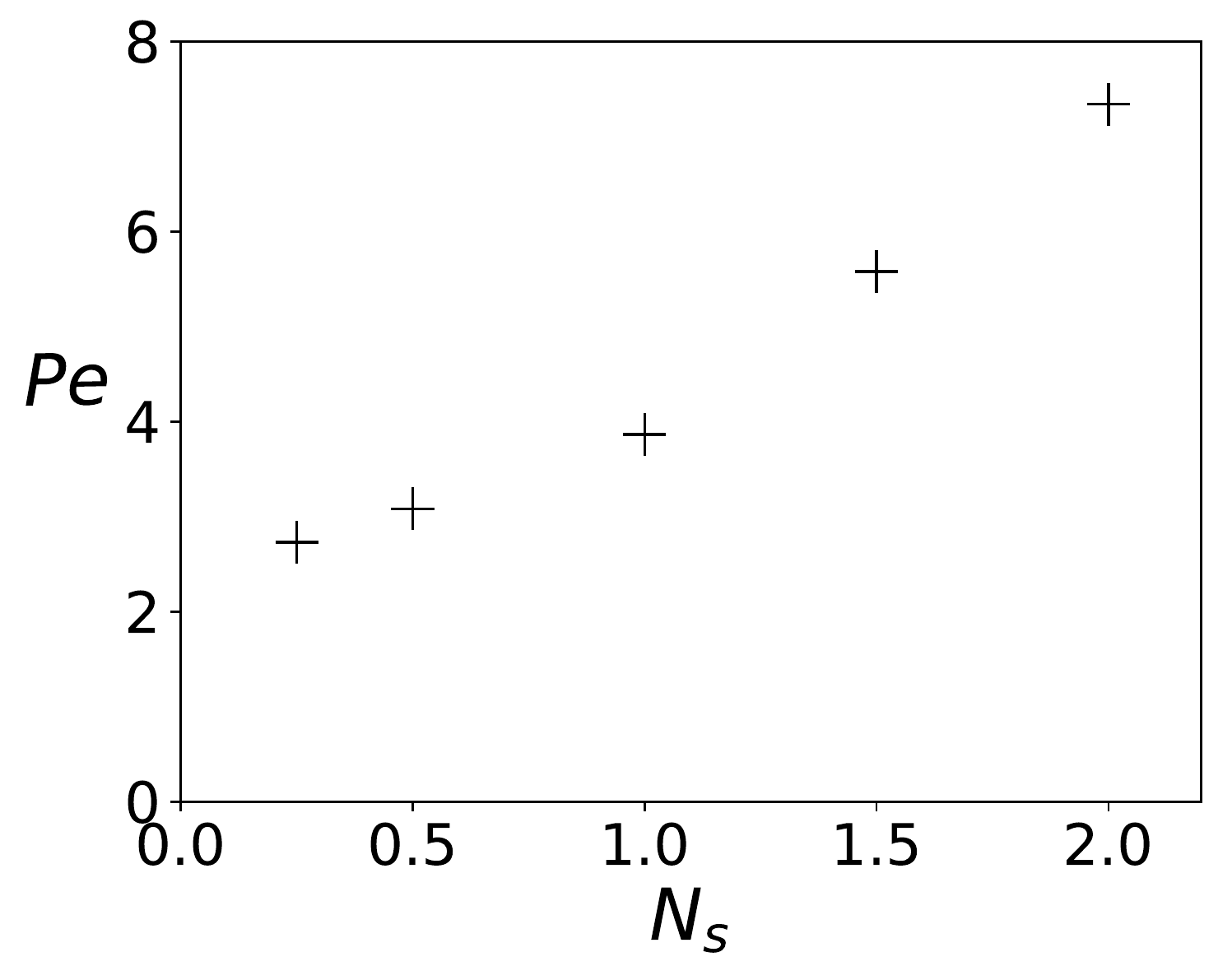}}
 \caption{(a) Travelling wave solution (\protect\tikz[baseline]{\protect\draw[line width=0.3mm] (0,.5ex)--++(0.65,0) ;}) and concentration profiles in the moving frame from DEM simulations at different times for the case $N_s=1$, $r=1.5$, $Pe = 3.86$. 
 $t=40435$ (\textcolor{C0}{\protect\tikz[baseline]{\protect\draw[line width=0.3mm,densely dashed] (0,.5ex)--++(0.65,0) ;}}), 
 $t=50543$ (\textcolor{C1}{\protect\tikz[baseline]{\protect\draw[line width=0.3mm,long dashdotted] (0,.5ex)--++(0.7,0) ;}}), 
 $t=60652$ (\textcolor{C2}{\protect\tikz[baseline]{\protect\draw[line width=0.3mm,dotted] (0,.5ex)--++(0.7,0) ;}}), 
 $t=70761s$ (\textcolor{C3}{\protect\tikz[baseline]{\protect\draw[line width=0.3mm,densely dash dot] (0,.5ex)--++(0.9,0) ;}}),
 $t=80870$ (\textcolor{C4}{\protect\tikz[baseline]{\protect\draw[line width=0.3mm,dash dot dot] (0,.5ex)--++(0.9,0) ;}}), (b) Value of the Peclet number as a function of the number of layers of small particles.}
 \label{fig:trw}
\end{figure}

For a travelling wave, the solution to equation~\eqref{eq:trwa} should not depend on time $\tau$. Assuming $\partial \phi_s/\partial \tau = 0$, this requires that $\tau D(\xi-c\ln(\tau))$ is independent of $\tau$. This implies
\begin{equation}
 \dfrac{\partial}{\partial \tau}\left(\tau D(\xi-c\ln(\tau))\right)=0,
\end{equation}
and distributing the derivative with $\tau$, the following differential equation is obtained
\begin{equation}
 D^{\prime}(\xi-c\ln(\tau)) -\dfrac{1}{c}D(\xi-c\ln(\tau)) =0,
\end{equation}
for which the solution is
\begin{equation}
 D = D_0e^{(\xi-c\ln(\tau))/c} = D_0e^{z/c}.
\end{equation}
This shows that the only way to obtain a time independent solution is that the diffusion coefficient has an exponential structure with the same vertical dependence as the advection flux $F_s$. In other words, the diffusion coefficient should also be proportional to the inertial number to the power $0.81$ in order to have a concentration profile of constant thickness. Under this assumption, equation~\eqref{eq:trwa} becomes
\begin{equation}
  c\dfrac{\partial \phi_s}{\partial \xi} - \dfrac{\partial}{\partial \xi}\left(S_{r0}e^{\xi/c}\phi_s(1-\phi_s)\right)
 = \dfrac{\partial}{\partial \xi}\left(D_0e^{\xi/c}\dfrac{\partial \phi_s}{\partial \xi} \right).
 \label{eq:trwa2}
\end{equation}
By integration, and imposing $\phi_s \rightarrow 0$ when $\xi \rightarrow \pm \infty$, the following equation is obtained
\begin{equation}
 \dfrac{\partial \phi_s}{\partial \xi} -\dfrac{S_{r0}}{D_0}\phi_s\left[ \phi_s-\left(1-\dfrac{c}{S_{r0}}e^{-\xi/c}\right)\right] = 0,
 \label{eq:trweq}
\end{equation}
for which the solution is
\begin{equation}
 \phi_s = \dfrac{e^{-\frac{S_{r0}}{D_0}\xi-\frac{c^2}{D_0}e^{-\xi/c}}}{C - \dfrac{S_{r0}}{D_0}\int_{-\infty}^{\xi}\left(e^{-\frac{S_{r0}}{D_0}\xi^{\prime}-\frac{c^2}{D_0}e^{-\xi^{\prime}/c}} \right)d\xi^{\prime}},
 \label{eq:trws}
\end{equation}
where $C$ is a constant satisfying the mass conservation $\int_{-\infty}^{+\infty} \phi_s d\xi$. The travelling wave solution has a complex dependency on $\xi$ and contains an integral which cannot be computed analytically. A semi-analytical approach is developed where the integral is computed numerically and a numerical optimization is performed on the $C$ coefficient so that $\int_{-\infty}^{+\infty} \phi_s d\xi$ corresponds to the initial mass of small particles introduced in the model. Figure~\ref{fig:trw} compares the travelling wave solution~\eqref{eq:trws} with concentration profiles from the DEM simulations (case $N_s=1$, $r=1.5$) at different times. The $D_0$ coefficient was computed in order to minimize the root mean square of the difference between DEM and the travelling wave solution. The travelling wave solution agrees very well with all DEM profiles. The width and the height of the profile are correctly predicted. However the latter is slightly below the DEM profiles. This may be explained by the uncertainties when computing the value of $S_{r0}$ from the DEM simulations. 

In order to keep the thickness of the small particles layer constant, both the advection and the diffusion coefficients must have the same exponential vertical structure. Defining a Peclet number by the ratio between the segregation intensity and the diffusion coefficient, $Pe = S_{r}(z)/D(z) = S_{r0}/D_0$, the travelling wave solution leads to a constant Peclet number with depth. The relative effect of segregation and diffusion fluxes is constant and independent of $z$. At each depth, the balance between advection and diffusion has to be the same in order to be consistent with the constant layer thickness observed in the DEM simulations. 

The value of the Peclet number depends on the number of layers of small particles $N_s$ and is plotted in figure~\ref{fig:trw}b. The value of $Pe$ increases almost linearly with $N_s$. Since the value of $S_{r0}$ is the same whatever the number of layers of small particles, this result suggests that the value of the diffusion coefficient decreases with $N_s$. This would indicate that diffusion is more efficient when there are fewer small particles. The mean free path of small particles is indeed expected to be larger if they are less concentrated.

This study has shown that both the segregation flux and diffusion need to be proportional to the inertial number to the power 0.81 in order to represent the dynamics of segregation in bedload transport.

\section{Conclusion}\label{sec:conclusion}
    Vertical size-segregation in bedload transport has been studied with a discrete and a continuous approach. Focusing on the quasi-static region - common to any granular flow on a pile - it has been shown with DEM simulations that small particles infiltrate the coarse quasi-static bed with the same behavior, ranging from a few isolated particles up to two layers of small particles. All the different configurations exhibit the same segregation velocity, related to a power law of the inertial number, generalizing the results observed for moderate inertial numbers in confined granular mixture without fluid \citep{fry2018}. The dynamics of the infiltration of small particles is totally controlled by the bottom of the small particle layer, as the inertial number decreases exponentially from the top to the bottom of the layer. As a consequence, the small particles segregate down as a layer of constant thickness. 

The continuous size-segregation model of \cite{gray2005}, \cite{thornton2006} and \cite{gray2006} has been shown to reproduce quantitatively the DEM simulations, with a small particle layer of constant thickness segregating at a velocity independent of the thickness of the layer.
In the continuous model, this comes from a perfect balance at any elevation between the effect of the inertial number and the local small particle concentration. This analysis demonstrates that the macroscopic segregation behavior always results from a local equilibrium between the inertial number forcing and the local small particles concentration. 
Moreover, based on the derivation of an analytical solution with a travelling wave approach, the results show that the diffusion coefficient and the segregation flux in the continuous model should have the same dependency on the inertial number.\\

This paper represents an original contribution on segregation processes in bedload transport, and more generally in dense granular flows. This is a first step toward upscaling grain size-segregation in continuous models for sediment transport. In the future, it would be interesting to use DEM simulations to infer the different grain-grain forces involved in size-segregation as well as collective effects related to particle concentration.

\section*{Acknowledgements}
This research was funded by the French ‘Agence nationale de la recherche’,
project ANR-16-CE01-0005 SegSed 'size segregation in sediment transport'.
The authors acknowledge the support of Irstea (formerly Cemagref).
This  research  was  supported  by  NERC  grants  NE/E003206/1  and  NE/K003011/1as  well  as  EPSRC  grants  EP/I019189/1,  EP/K00428X/1  and  EP/M022447/1. J.M.N.T.G.  is  a  Royal  Society  Wolfson  ResearchMerit   Award   holder   (WM150058)   and   an   EPSRC   Established   Career   Fellow(EP/M022447/1).

\appendix
\section{Solution of the purely advective segregation model}\label{appA}

\cite{may2010} already derived a solution to the model of \cite{thornton2006} with an exponential dependence of the segregation flux with $z$. In this appendix, the full solution is presented in the bedload configuration. The segregation equation to solve is:

\begin{equation}
\dfrac{\partial \phi_s}{\partial t} - \dfrac{\partial F_s}{\partial z} = 0,
\label{eq:hypeq}
\end{equation}
with the flux $F_s = S_{r0}e^{z/c}\phi_s(1-\phi_s)$ and an initial step condition in concentration:
\begin{equation}
 \phi_{s0} = \phi_s(z, 0) =
 \left\{
 \begin{array}{lr}
  0, & \quad z < z_i,\\
  1, & \quad z \geq z_i.
 \end{array}
 \right.
 \label{eq:sol_init_ann}
\end{equation}
The boundary conditions ensure that there is no flux at $z=0, H$, which requires that
\begin{equation}
  \phi_s = 0 \text{ or } 1, \quad \text{at} \quad z=0 \text{ and }H.
  \label{eq:bound_cond}
 \end{equation}
Substituting the flux into~\eqref{eq:hypeq} implies
\begin{equation}
 \dfrac{\partial \phi_s}{\partial t} + W(z, \phi_s)\dfrac{\partial
  \phi_s}{\partial z} = S(z, \phi_s)
  \label{eq:edv_eq}
\end{equation}
where
\begin{align}
\label{eq:W}
 W(z, \phi_s) &= S_{r0}e^{z/c}\left(2\phi_s -1\right),\\
 \label{eq:S}
 S(z, \phi_s) &= \dfrac{S_{r0}}{c}e^{z/c}\phi_s(1-\phi_s).
\end{align}
Due to dependence of the flux both on $z$ and $\phi_s$, the equation to solve has a non-linear advective velocity and a source term. In the following, this problem is solved using the method of characteristics.

On a characteristic curve $\left(Z(s), t(s)\right)$, 
\begin{equation}
 \dfrac{d\phi_s}{ds} = \dfrac{\partial \phi_s}{\partial t}\dfrac{dt}{ds} + \dfrac{\partial \phi_s}{\partial z}\dfrac{dz}{ds} = S.
 \label{eq:charac_curve}
\end{equation}
Comparing coefficients of~\eqref{eq:edv_eq} and ~\eqref{eq:charac_curve}
\begin{align}
 \dfrac{d\phi_s}{ds}&=S,\\
 \dfrac{dt}{ds}&=1,\\
 \dfrac{dZ}{ds}&=W.
\end{align}
Eliminating $s$ implies 
\begin{align}
 \dfrac{d\phi_s}{dt}&=S\\
 \dfrac{dZ}{dt}&=W.
\end{align}

%

The problem~\eqref{eq:edv_eq} is thus equivalent to the following system:
\begin{equation}
\left\{
 \begin{array}{rcl}
   \dfrac{d}{dt}\phi_s(Z(t),t) &=& S\left(Z(t), \phi_s\left(Z(t),t\right)\right),\\
   \\[.5pt]
   \dfrac{dZ(t)}{dt} &=& W\left(Z(t), \phi_s\left(Z(t),t\right)\right),\\
   \\[.5pt]
    Z(0) &=& z_0.
  \end{array}
  \right.
  \label{eq:cauchy_equiv}
\end{equation}

The first equation indicates that $\phi_s$ is not constant along a characteristic curve and the second equation that the characteristic curves are not linear.
The method of characteristics consists in solving the value of $\phi_s$ along the characteristics curves. If a characteristic curve passing by each time and space position exists, then the problem can be fully resolved. In order to compute the value of $\phi_s$ along a characteristic curve, the first equation of \eqref{eq:cauchy_equiv} is differentiated with time
\begin{equation}
 \dfrac{d^2}{dt^2}\phi_s(Z(t),t) = \dfrac{d}{dt} \left[ S\left(Z(t), \phi_s\left(Z(t),t\right)\right) \right],
\end{equation}
distributing the derivative,
\begin{equation}
 \dfrac{d^2}{dt^2}\phi_s(Z(t),t) = \dfrac{dZ}{dt}\dfrac{\partial S}{\partial z} + \dfrac{d\phi_s\left(Z(t),t\right)}{dt}\dfrac{\partial S}{\partial \phi_s},
\end{equation}
and introducing the first and second equation of~\eqref{eq:cauchy_equiv} it implies,
\begin{equation}
 \dfrac{d^2}{dt^2}\phi_s(Z(t),t) = W\dfrac{\partial S}{\partial z} + S\dfrac{\partial S}{\partial \phi_s}.
 \label{eq:derivphi}
\end{equation}
Using equation~\eqref{eq:W} and \eqref{eq:S}, one can remark that $\partial S/\partial z = S/c$ and $\partial S/\partial \phi_s = -W/c$. Introducing this in~\eqref{eq:derivphi}, the second derivation with time of $\phi_s$ along a characteristic curve is identically null,
\begin{equation}
 \dfrac{d^2}{dt^2}\phi_s(Z(t),t)  = 0.
\end{equation}
Integrating twice with time, a linear dependence on time of $\phi_s$ is obtained,
\begin{equation}
\phi_s(Z(t),t) = \phi_{s0}(z_0) + S\left(z_0, \phi_{s0}(z_0)\right)t.
\label{eq:phi_along2}
\end{equation}
The value of $\phi_s$ along the characteristic curve is entirely determined by its inititial value and therefore by the initial condition $\phi_{s0}$. It is now possible to compute the position with time of a characteristic curve. Introducing \eqref{eq:phi_along2} in the second equation of \eqref{eq:cauchy_equiv} and integrating, a relation for the characteristic curves is obtained,
\begin{align}
\label{eq:characteristic1}
 & e^{-Z(t)/c} - e^{-z_0/c} = -\dfrac{S_{r0}}{c}(2\phi_{s0}(z_0)-1)t - \dfrac{S_{r0}}{c}S\left(z_0, \phi_{s0}(z_0)\right)t^2,\\
 \label{eq:characteristic2}
 \Leftrightarrow \quad & Z(t) = z_0 -c\ln \left(1 -\dfrac{1}{c}W\left(z_0, \phi_{s0}(z_0)\right)t - \dfrac{S_{r0}}{c}e^{z_0/c}S\left(z_0, \phi_{s0}(z_0)\right)t^2\right),
\end{align}
and the characteristic curve is entirely determined by the value of the initial solution $\phi_{s0}$ in its origin position $z_0$. 
Finally the solution to system \eqref{eq:cauchy_equiv} is
\begin{equation}
\left\{
 \begin{array}{rcl}
    \phi_s(Z(t),t) &=& \phi_{s0}(z_0) + S\left(z_0, \phi_{s0}(z_0)\right)t,\\
    \\[.5pt]
    Z(t) &=& z_0 -c\ln \left(1 -\dfrac{1}{c}W\left(z_0, \phi_{s0}(z_0)\right)t - \dfrac{S_{r0}}{c}e^{z_0/c}S\left(z_0, \phi_{s0}(z_0)\right)t^2\right),\\
  \end{array}
  \right.
  \label{eq:charac_solution}
\end{equation}

\begin{figure}
 \centering
 \subfloat[]{\includegraphics[scale=0.4]{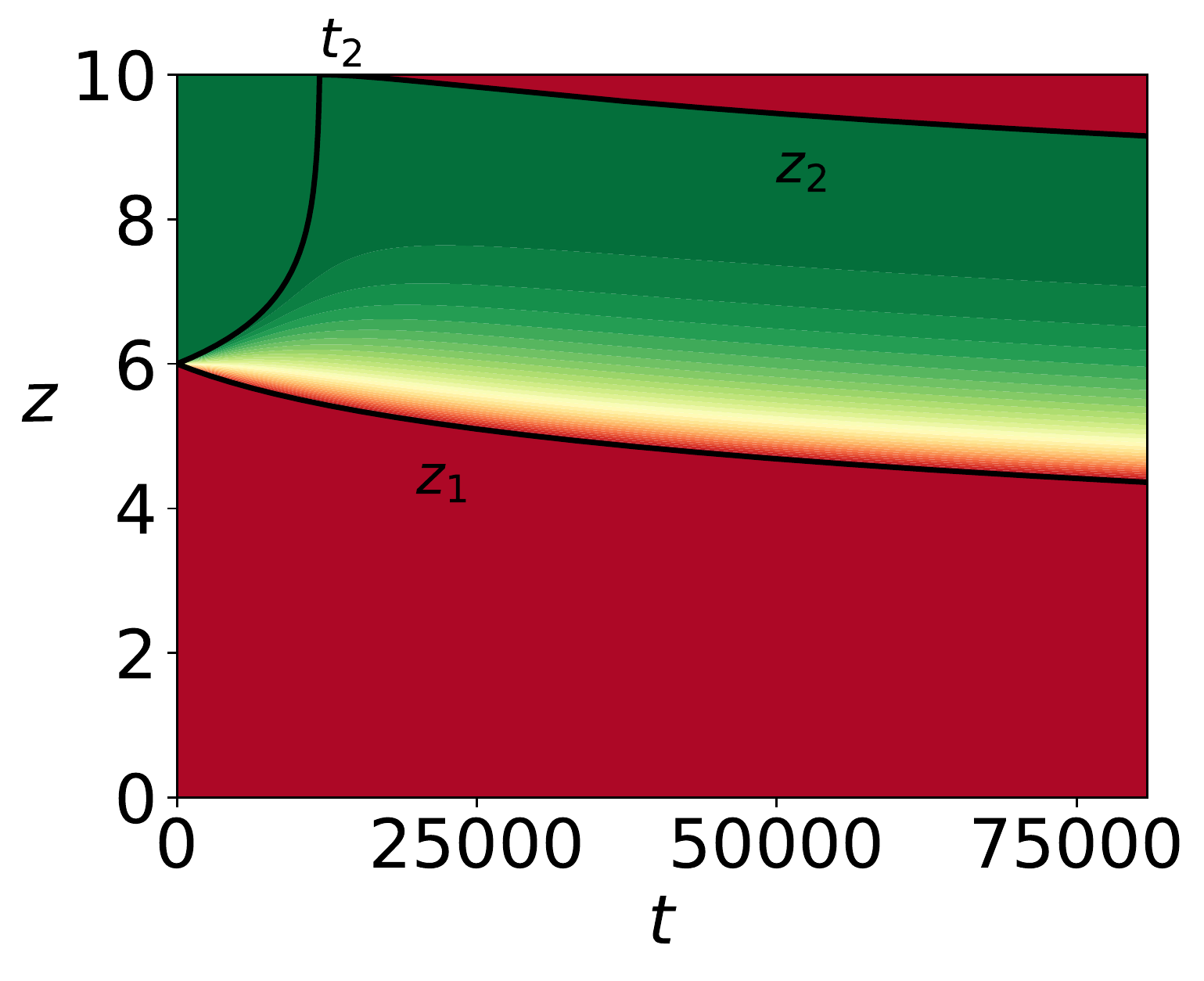}}
 \subfloat[]{\includegraphics[scale=0.4]{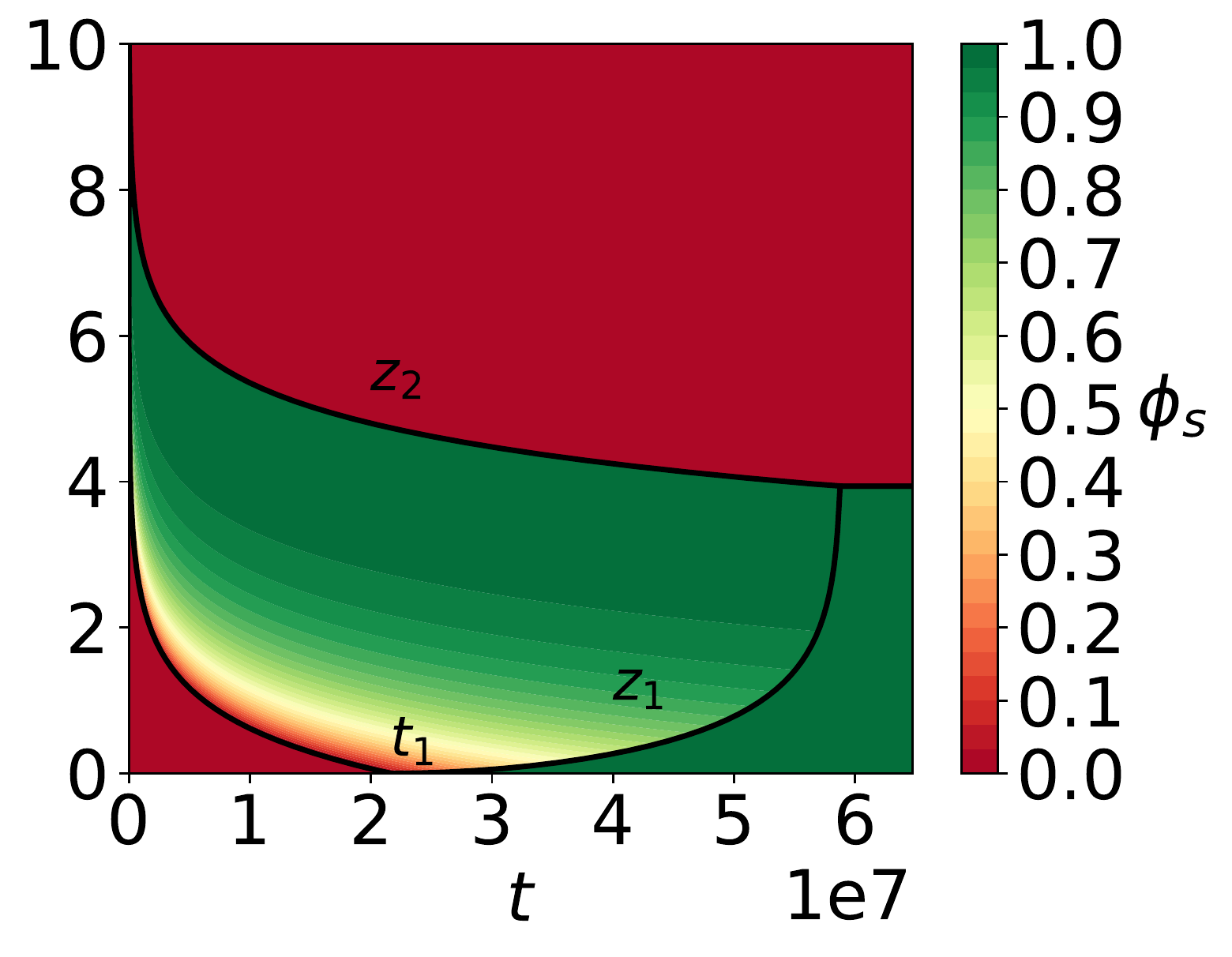}}
 \caption{Example of solution obtained by the method of characteristics. $z_i = 6$ (a) First time steps, (b) full solution.}
 \label{fig:solution_example}
\end{figure}

Equation \eqref{eq:charac_solution} is a general solution for any initial condition. In the present configuration the initial solution \eqref{eq:sol_init_ann} is a step of concentration with a discontinuity at $z_i$ and represents an initial state where a pure layer of small particles is placed above a bed of only large particles. Figure~\ref{fig:solution_example} shows the full solution for the case $z_i = 6$, and the derivation of this solution is presented below. According to the second equation of~\eqref{eq:cauchy_equiv}, the characteristic curves coming from $z_0 < z_i$ have a negative velocity and are going down. Conversely, those coming from $z_0 > z_i$ are going up. Therefore there is a rarefaction fan, where no characteristic curves are present.

For $z_0 < z_i$, $\phi_{s0}(z_0)=0$, the equation of the characteristic curves are $Z(t) = z_0 - c\ln\left(1+\frac{S_{r0}}{c}e^{z_0/c}t\right)$. Along these characteristics, $\phi_s(Z(t),t) = 0$. $z_1$ is defined as the characteristic curve at the bottom of the rarefaction zone (coming from $z_i$): $z_1(t) = z_i - c\ln\left(1+\frac{S_{r0}}{c}e^{z_i/c}t\right)$. $z_1$ is the lower bound of the rarefaction zone. The time at which $z_1$ meets the bottom of the domain is called $t_1 = \frac{c}{S_{r0}}(1-e^{-z_i/c})$ and corresponds to the time when the first small particle reaches the bottom.

For $z_0 > z_i$, $\phi_{s0}(z_0)=1$, the equation of the characteristic curves are $Z(t) = z_0 - c\ln\left(1-\frac{S_{r0}}{c}e^{z_0/c}t\right)$. Along these characteristics, $\phi_{s}(Z(t),t) = 1$. $z_2$ is defined as the characteristic curve coming from $z_i$: $z_2(t) = z_i - c\ln\left(1-\frac{S_{r0}}{c}e^{z_i/c}t\right)$. $z_2$ is the upper bound of the rarefaction zone. The time at which $z_2$ meets the top of the domain is called $t_2 = \frac{c}{S_{r0}}(e^{-z_i/c}-e^{-H/c})$ and corresponds to the time when the first large particle reaches the top.

Once $z_2$ reaches the top of domain at $t_2$ the boundary condition~\eqref{eq:bound_cond} implies that $\phi(H,t_2)$ jumps from $1$
to $0$. A shock forms and the position of the shock must satisfy the Rankine-Hugoniot relation, which ensures the conservation of momentum through the shock :
$$
z_2^{\prime} = -\dfrac{F_s(\phi^+) -F_s(\phi^-)}{\phi^+ - \phi^-}.
$$	
Above the shock $\phi^+ = 0$ and below $\phi^-$ is the solution in the rarefaction zone:
\begin{equation}
 z_2^{\prime} = -S_{r0}e^{z_2/c}\left(1-\phi_s(z_2,t)\right).
 \label{eq:shock_equation}
\end{equation}

A similar expression applying the Rankine-Hugoniot relation is found for $z_1$ for $t>t_1$. For now, as no solution in the rarefaction has been computed, the shock evolution equation~\eqref{eq:shock_equation} cannot be solved. The solution in the rarefaction zone is now computed. The rarefaction zone is delimited by $z_1$ and $z_2$. Due to the discontinuity in the initial solution, no characteristic curves are present in the rarefaction wave ($z_1 \leq z \leq z_2$). In order to find a solution, it is considered that there is an infinity of characteristic curves coming from the point of discontinuity $z_i$, each of them being associated to a different initial value, noted $\bar{\phi}_{s0}$ and verifying $\phi_{s0}(z_c)^- = 0\leq \bar{\phi}_{s0} \leq 1 = \phi_{s0}(z_c)^+$. In other words, for $(z,t)$ in the rarefaction zone, one looks for a characteristic curve coming from $z_i$ associated to an initial value $\bar{\phi}_{s0}$ and verifying system~\eqref{eq:charac_solution}:
\begin{equation}
 \left\{
 \begin{array}{rcl}
  \phi_s(z,t) &=& \bar{\phi}_{s0} + S\left(z_i, \bar{\phi}_{s0}\right)t,\\
  e^{-z/c} - e^{-z_i/c} &=& -\dfrac{S_{r0}}{c}(2\bar{\phi}_{s0}-1)t - \dfrac{S_{r0}}{c}e^{z_i/c}S\left(z_i, \bar{\phi}_{s0}\right)t^2.
 \end{array}
 \right.
 \label{eq:rarefaction_sol}
\end{equation}

To compute the value of $\phi_s$ in the rarefaction zone with the first equation of~\eqref{eq:rarefaction_sol}, only the value of $\bar{\phi}_{s0}$ is missing. It is computed by an invertion of the second equation of \eqref{eq:rarefaction_sol} (quadratic equation and choosing the relevant solution):
\begin{equation}
 \bar{\phi}_{s0} = \dfrac{1}{2} + \dfrac{c}{S_{r0}t}e^{-z_i/c}\left(1 - e^{(z_i-z)/c}\sqrt{1+\left(\dfrac{S_{r0}t}{2c}\right)^2e^{(z+z_i)/c}} \right).
 \label{eq:phi0_bar}
\end{equation}

Inserting~\eqref{eq:phi0_bar} into the first equation of \eqref{eq:rarefaction_sol} the solution in the rarefaction zone is:
\begin{equation}
 \phi_s(z,t) = \dfrac{1}{2} - \dfrac{c}{S_{r0}t}e^{-z/c} + \dfrac{c}{S_{r0}t}e^{-(z+z_i)/(2c)}\sqrt{1+\left(\dfrac{S_{r0}t}{2c}\right)^2e^{(z+z_i)/c}}.
 \label{eq:phis_sol}
\end{equation}

The equation of the shock~\eqref{eq:shock_equation} can now be solved. However due to the complex form of the solution in the rarefaction zone~\eqref{eq:phis_sol} the equation of the shock~\eqref{eq:shock_equation} is solved numerically. Finally the full solution is presented on figure~\ref{fig:solution_example}.

\bibliographystyle{natbib}
\bibliography{biblio}

\end{document}